\documentclass[reprint,aps,pra,superscriptaddress,noeprint]{revtex4-2}

\usepackage{bm}
\usepackage{amsthm}
\usepackage{amsmath,amssymb}
\usepackage{graphicx}
\usepackage{braket}
\usepackage{enumerate}
\usepackage{ascmac}
\usepackage{mathrsfs}

\usepackage[
colorlinks=true,
filecolor=blue,
urlcolor=blue
]{hyperref}
\usepackage{mathtools,bbding}
\usepackage{CJKutf8}
\usepackage{qcircuit}
\usepackage{subcaption}

\theoremstyle{definition}
\newtheorem{theorem}{Theorem}

\newtheorem{definition}[theorem]{Definition}
\newtheorem{lemma}{Lemma}
\newtheorem{corollary}[theorem]{Corollary}
\newtheorem{example}[theorem]{Example}
\newtheorem{remark}[theorem]{Remark}
\newtheorem{claim}[theorem]{Claim}

\newcommand{\pd}[3]{
 \if 1#1 \frac{\partial #2}{\partial #3}
 \else \frac{\partial^{#1} #2}{\partial #3^{#1}}\fi}
 \newcommand{\od}[3]{
 \if 1#1 \frac{{\mathrm d} #2}{{\mathrm d} #3}
 \else \frac{{\mathrm d}^{#1} #2}{{\mathrm d}#3^{#1}}\fi}

\newcommand{\e}[1]{\begin{align} #1\end{align}}
\newcommand{\h}[1]{\hat{#1}}
\newcommand{\tr}[1]{\mathrm{tr}\left[ #1 \right]}

\newcommand{\ex}[0]{{\rm e}}
\newcommand{\expo}[1]{\exp{\left[ #1 \right]}}
\newcommand{\calD}[0]{ {\mathcal D}}
\newcommand{\calE}[0]{ {\mathcal E}}
\newcommand{\calH}[0]{{\mathcal H}}
\newcommand{\calI}[0]{{\mathcal I}}

\newcommand{\calM}[0]{{\mathcal M}}
\newcommand{\calN}[0]{{\mathcal N}}
\newcommand{\vac}[0]{{\rm vac}}

\newcommand{\kakko}[1]{\left( #1 \right)}
\newcommand{\intd}[0]{{\rm d}}

\DeclareMathOperator{\re}{Re}
\DeclareMathOperator{\im}{Im}

\begin{document}

\title{Exploiting Translational Symmetry for Quantum Computing \\
with Squeezed Cat Qubits
}

\author{Tomohiro Shitara}
\email{tomohiro.shitara@ntt.com}
\affiliation{NTT Computer and Data Science Laboratories, NTT Corporation, Musashino, 180-8585, Tokyo, Japan} 

\author{Gabriel Mintzer}
\affiliation{MIT Department of Electrical Engineering and Computer Science, 50 Vassar St, Cambridge, MA 02139, USA}

\author{Yuuki Tokunaga}
\affiliation{NTT Computer and Data Science Laboratories, NTT Corporation, Musashino, 180-8585, Tokyo, Japan} 

\author{Suguru Endo}
\email{suguru.endou@ntt.com}
\affiliation{NTT Computer and Data Science Laboratories, NTT Corporation, Musashino, 180-8585, Tokyo, Japan}

\date{\today}

\begin{abstract}
Translational symmetry plays an essential role in bosonic quantum error correction (QEC), most notably in the Gottesman-Kitaev-Preskill code. Squeezed cat (SC) codes provide a complementary platform, combining approximate protection against physical errors with the noise bias of cat codes, but a hardware-efficient route to exploit their translational symmetry for QEC has been lacking. Here we show that this symmetry provides a practical route to autonomous QEC and universal quantum computation with SC codes. We then propose a QEC protocol that autonomously restores states driven out of the code space by physical errors, even though translational symmetry along a single direction does not uniquely define the code space. Using a subsystem decomposition based on squeezed displaced Fock states, we analytically characterize the relaxation rate toward the code space induced by the protocol, thereby estimating the QEC-cycle rate required for effective error suppression. Within the same framework, we propose deterministic preparation of logical states, logical gates, and logical-$Z$ readout with improved error scaling. These results establish translational symmetry as a new perspective for approaching quantum computation with SC qubits.

\end{abstract}

\maketitle

\textit{\textbf{Introduction}}---
 Quantum error correction (QEC) codes play a crucial role in addressing the challenge of noise corrupting quantum states~\cite{Shor1995, Bennett1996a,Laflamme1996,Knill1996,Nielsen2000}. A key drawback of qubit-based QEC codes is that they require entanglement among many physical qubits to robustly encode logical qubits~\cite{Bennett1996a,Laflamme1996,Knill1996}. This has motivated the development of an alternative type of QEC codes known as bosonic QEC codes~\cite{Cochrane1999,Gottesman2001,Michael2016,Grimsmo2020,Walshe2020,Noh2020,Royer2022}. Bosonic QEC codes generally utilize only a very small number of continuous-variable modes because they encode information in the infinite-dimensional Hilbert space of a single bosonic mode.
 
Bosonic QEC codes are generally classified according to their symmetries, either rotational or translational.  Rotation-symmetric codes~\cite{Grimsmo2020} are exemplified by cat codes~\cite{Cochrane1999, Mirrahimi2014}. For example, the two-legged cat code is a variety of rotation-symmetric code with an error-biased code property---i.e., it is very robust to phase errors but susceptible to photon-loss errors, having no capability to correct the latter. Meanwhile, Gottesman-Kitaev-Preskill (GKP) codes are characterized by translational symmetries, and they offer a hardware-friendly QEC protocol implemented by imitating the dissipative process with the help of an ancilla qubit~\cite{Royer2020}. However, the orthogonality of the GKP codewords is restricted by the experimentally-realizable squeezing level, and suboptimal orthogonality induces inevitable readout errors. 

Recently, squeezed cat (SC) codes have emerged as practical bosonic QEC codes~\cite{Hillmann2023,Schlegel2022}. SC codes are squeezed variants of the biased cat codes, and the translational symmetry of the SC code states is more evident than that of the cat code states. Note that the SC codes not only inherit the biased-noise property of cat codes but also further suppress the logical error rate due to photon-loss error by optimizing the amplitude and squeezing level~\cite{Schlegel2022,Xu2023}.
Furthermore, the photon-loss errors cause the state to leak from the SC code space, allowing for the partial detection and correction of logical errors. However, the dissipative QEC strategies proposed thus far either cannot suppress the logical error~\cite{Schlegel2022, Hillmann2023}, as we will show later, or require experimentally demanding nonlinear interaction among ancillary systems~\cite{Xu2023}.

\begin{figure}[htbp]
    \centering
    \includegraphics[width=\linewidth]{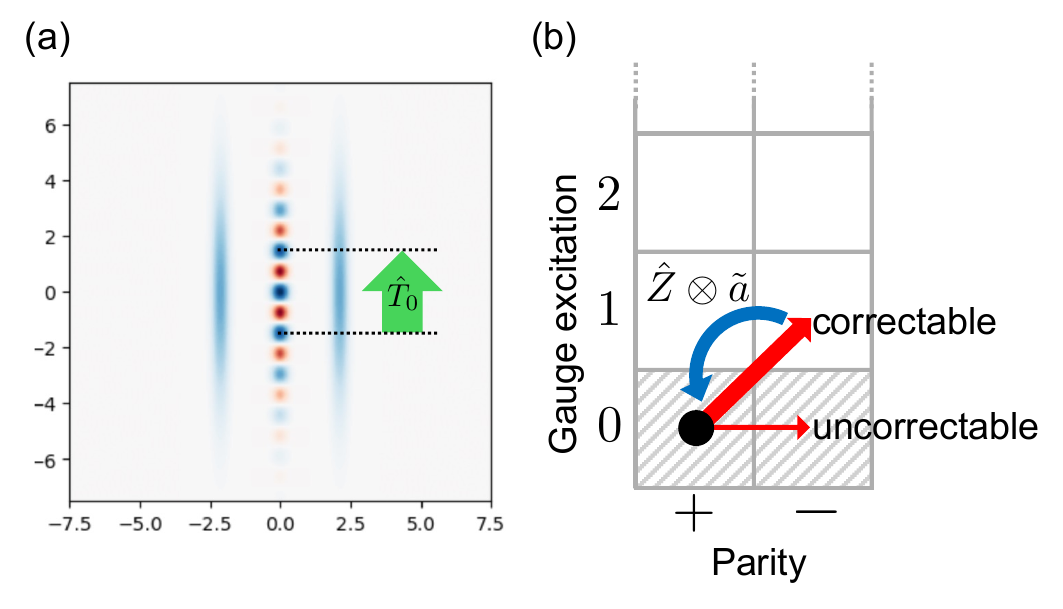}
    \caption{(a) Wigner function of the SC state $\ket{{\rm sq}^+_{\alpha, r}}$ (Eq.~\eqref{eq:sq}) and its translational symmetry $\h T_0$.
    (b) Subsystem decomposition of the bosonic Hilbert space. The photon-loss process (red arrows) changes the parity and partially generates an excitation in the gauge space. 
    Our proposed QEC protocol (blue arrow) approximately dissipates back to the SC code space (shaded area), along with the parity change, and thus appropriately corrects the correctable part of the photon-loss error.
    }
    \label{fig:schematic}
\end{figure}

In this letter, we unravel the unexplored utility of the translational symmetry and its underlying mathematical structure in SC codes for quantum computation by proposing protocols for hardware-efficient autonomous QEC, robust logical operations, and precise readout.
We first show that SC states are exactly stabilized by a non-unitary operator associated with the incomplete translational symmetry due to finite squeezing.
Building on this observation, we construct a corresponding dissipator and design an autonomous QEC protocol, i.e., QEC without syndrome measurements and feedback operations.
Based on an analysis using a recently introduced subsystem decomposition with the squeezed displaced Fock states~\cite{Putterman2022, Chamberland2022, Xu2023}, we show that the proposed protocol possesses two key features: (i) although a single-directional symmetry does not uniquely define the SC code space, a sufficiently large squeezing level ensures effective dissipative stabilization into the code space; and (ii) the protocol corrects logical errors induced by photon loss.
Our QEC protocol is hardware-efficient because it can be performed by the repeated alternation of a conditional displacement operation involving the ancilla qubit and the resonator with an ancilla-qubit reset.
We then show that high-fidelity logical operations become available by interspersing QEC operations throughout lower-fidelity logical operations.
Finally, we introduce an efficient and precise measurement method for the logical $Z$ operator constructed from the non-unitary stabilizer.

\textit{\textbf{Preliminaries}}---
We start from the displaced squeezed state defined by
$\ket{\alpha, z} := \h{D}(\alpha) \h{S}(z)\ket{\vac}$,
where $\h{D}(\alpha) := e^{\alpha \h{a}^\dagger - \alpha^* \h{a}}$ and $\h{S}(z) := e^{\frac{1}{2}(z^* \h{a}^2 - z \h{a}^{\dagger 2})}$ are the displacement operator and the squeezing operator, respectively. In terms of this state, the SC state is defined as
\e{
\ket{{\rm sq}^\pm_{\alpha, r}} \coloneq \frac{1}{\sqrt{\calN_{0}^{\pm}}}\kakko{\ket{\alpha, r} \pm \ket{-\alpha, r}}, \label{eq:sq}
}
where 
$\calN_{0}^{\pm}=\kakko{\bra{\alpha, r} \pm \bra{-\alpha, r}}\kakko{\ket{\alpha, r}\pm \ket{-\alpha, r}}$
is the normalization constant. We assume that the displacement amplitude $\alpha$ and the squeezing parameter $r$ are both real and positive throughout the paper.
Analysis based on the Knill-Laflamme conditions predicts that the SC codes potentially have resilience against both photon-loss and dephasing errors~\cite{Schlegel2022}.
For example, while photon-loss errors are not detectable in the conventional cat codes, they \emph{can} be partially detected and corrected in the SC codes, as shown later using subsystem decomposition. 

The SC states possess an approximate discrete translational symmetry~\cite{Schlegel2022, Endo2024}, as shown in Fig.~\ref{fig:schematic} (a). Indeed, they satisfy
\e{
\bra{{\rm sq}^\pm_{\alpha,    r}} \h{D}(i \xi) \ket{{\rm sq}^\pm_{\alpha,    r}} =  \expo{-\frac{1}{2}e^{-2 r} \xi^2} \cos (2\alpha \xi),
}
indicating a period of $\frac{\pi}{\alpha}$ with respect to $\xi \in \mathbb{R}$ if we neglect the exponentially decaying prefactor. In the infinite-squeezing limit of $r \to\infty$, the approximate symmetry becomes exact as $\h{T}_0 \ket{{\rm sq}^\pm_{\alpha,    r = \infty}} = \ket{{\rm sq}^\pm_{\alpha,    r = \infty}}$, where $\h{T}_0 = \h{D}\left( \frac{i \pi}{\alpha} \right)$ is a stabilizer operator.
We will explicitly leverage this translational symmetry of the SC states for our QEC protocol.

As a convenient tool for theoretical analyses, we introduce the subsystem decomposition of the bosonic Hilbert space~\cite{Putterman2022,Chamberland2022,Xu2022,Xu2023}, as well as a natural basis on it.
Noting that the SC states are generated from the vacuum state followed by displacement and squeezing, we define the squeezed displaced Fock (SDF) state as
\e{
\ket{\Psi_n^\pm} = \frac{1}{\sqrt{\calN_n^\pm}}\h S(r)\kakko{\h{D}(\alpha') \pm (-1)^n \h{D}(-\alpha')}  \ket n,
}
where $\calN_n^\pm$ is the normalization constant and $\alpha' = \alpha  e^r$ is the rescaled displacement.
The sign $\pm$ corresponds to the parity of the photon number, i.e., $ e^{i\pi \h{a}^\dagger \h{a}} \ket{\Psi_n^\pm} = \pm \ket{\Psi_n^\pm}$, defining the orthogonal parity sectors as $\braket{\Psi_n^+|\Psi_m^-} = 0$.
However, the states of the same parity are non-orthogonal, with their overlap being $O( e^{-2 \alpha'^2})$.
To construct an orthonormal basis, we perform the Gram-Schmidt orthonormalization procedure from $n = 0$ in each parity sector. As such, we obtain the complete orthonormal set on the bosonic Hilbert space denoted by
$\ket{\pm}_L \otimes \ket{\tilde n}_G \simeq \ket{\Psi_n^\pm}$, which is called the SDF basis~\cite{Xu2023}.
The subscript $L$ ($G$) represents the logical (gauge) degree of freedom, and the SC codewords correspond to the ground state in the gauge mode: $\ket{{\rm sq}^\pm_{\alpha,    r}} = \ket{\pm}_L \otimes \ket{\tilde 0}_G$.
The total Hilbert space $\calH$ is then decomposed as $\calH \simeq \calH_L \otimes \calH_{G}$, with $\dim\calH_L = 2$ and $\dim\calH_G = \infty$; this decomposition is termed the \textit{subsystem decomposition}.

The SDF basis provides a powerful tool for analyzing errors described by the annihilation and creation operators. For example, in this basis, the annihilation operator is expressed as~$\h{a} \simeq  \h{Z}_L \otimes \kakko{\tilde{a} \cosh{r} - \tilde{a}^\dagger \sinh{r} +\alpha \tilde{I}}$~\cite{Xu2023}.
Here, $\h{Z}_L$ is the logical Pauli $Z$ operator acting on the logical space, and $\tilde{a}, \tilde{a}^\dagger, \tilde{I}$ are the annihilation, creation, and identity operators acting on the gauge space, respectively.
This expression has profound implications for the correctability of the photon-loss errors of the squeezed cat code.
First, the photon-loss error always induces a logical phase-flip error $\h Z_L$, since the single photon-loss process changes the parity in the bosonic mode.
Second, the gauge mode is partially modified at the same time, which can be used as a syndrome to correct the logical error $\h Z_L$ (see Supplemental Material (SM) for details
\footnote{See Supplemental Material.
}).

\textit{\textbf{Stabilizer and dissipator}}---
To derive the autonomous QEC protocol, we first identify the non-unitary symmetry operator that exactly stabilizes the SC state $\ket{{\rm sq}^\pm_{\alpha, r}}$.
As shown in SM~\cite{Note1}, the finitely-squeezed SC state is obtained by applying the envelope operator $\h{E}_\Delta =  e^{-\Delta^2 \h{p}^2 / 2}$ to the infinitely-squeezed cat state as $\h{E}_\Delta \ket{{\rm sq}^\pm_{\alpha,    r = \infty}}\propto \ket{{\rm sq}^\pm_{\alpha,    r}}$.
Here, the parameter $\Delta$ determines the width of the envelope and is related to the squeezing parameter $r$ by $\Delta =  e^{-r}$.
Then, the stabilizer operator is modified as
\e{
\h{T}_0 \to \h{T}_\Delta = \h{E}_\Delta \h{T}_0 \h{E}_\Delta^{-1} = e^{\frac{\sqrt{2} \pi}{\alpha}(i \h{x} - \Delta^2 \h{p})},
\label{eq:stab}
}
which is a non-unitary operator. We define the dissipator $\h d_\Delta$ by~\footnote{The dissipator is chosen so that $\h{d}_\Delta \propto \log \h{T}_\Delta$ and it is normalized in the sense that it behaves like an annihilation operator if we neglect the modularity---that is, $\lim_{\alpha \to \infty}[\h{d}_\Delta, \h{d}_\Delta^\dagger] = \h{I}$.}
\e{
\h d_\Delta=-\frac{i\alpha}{2\pi\Delta}\log \h T_\Delta=\frac{1}{\sqrt{2}} \kakko{\frac{\h{x}_{[\sqrt{2} \alpha]}}{\Delta} + i \Delta \h{p}},
}
where $\h{x}_{[\sqrt{2} \alpha]}$ is the modular position operator~\cite{Zak1967,Pantaleoni2023,Note1} satisfying $\h{x}_{[\sqrt{2} \alpha]} = \h{x} \ {\rm mod}\ \sqrt{2} \alpha$ and $-\alpha / \sqrt{2} < \h{x}_{[\sqrt{2} \alpha]} \le \alpha / \sqrt{2}$. 
Noting that the stability condition for a state is equivalent to annihilating it by the dissipator, i.e., $\h{T}_\Delta \ket{\psi} = \ket{\psi} \Leftrightarrow \h{d}_\Delta \ket{\psi} = 0$, it follows that we can stabilize the squeezed cat states by the dissipative process with the dissipator $\h d_\Delta$.

One of our main findings is that, in the SDF basis, the dissipator $\h d_\Delta$ can be expressed as~\cite{Note1}
\e{
\h{d}_\Delta = \h{Z}_L \otimes \frac{\tilde{x}_{[\sqrt{2} \alpha']} + i \tilde{p}}{\sqrt{2}},
}
where $\tilde x_{[\sqrt{2} \alpha']}$ and $\tilde p$ are the modular position operator and momentum operator acting on the gauge mode, respectively.
In the limit $\alpha' =  e^r \alpha \rightarrow \infty$, the period of the modular position operator diverges and the modularity becomes effectively negligible, so we have
\e{
\h{d}_\Delta \simeq \h{Z}_L \otimes \tilde{a} \quad (\alpha' \to \infty),  \label{eq:dissipator}
}
which coincides with the expression obtained in Ref.~\cite{Xu2023}. 
This dissipator can correct the correctable part of the logical error due to photon loss. 
Note that other proposed autonomous QEC methods~\cite{Schlegel2022,Hillmann2023} cannot correct logical errors, since the dissipator is quadratic in $\hat{a}$ and $\h a^\dagger$ and hence has even parity.

It is worth noting that our dissipator stabilizes in only one direction and that the steady-state space is strictly larger than the SC code space, in contrast with the cases for the SC states in Ref.~\cite{Xu2023} and the GKP states in Ref.~\cite{Royer2020}, where the dissipator is designed so that the steady-state space coincides with the code space.
Nevertheless, in the limit of $\alpha' \to \infty$, the periodicity in the modular operator becomes effectively negligible and $\h{d}_\Delta$ dissipates the gauge mode to the vacuum state.
We also note that the limit in Eq.~\eqref{eq:dissipator} is a state-dependent notion, i.e., a weak limit, which limits the applicability of our QEC protocol, as discussed later.

\begin{figure}[htbp]
    \centering
    \includegraphics[width=\linewidth]{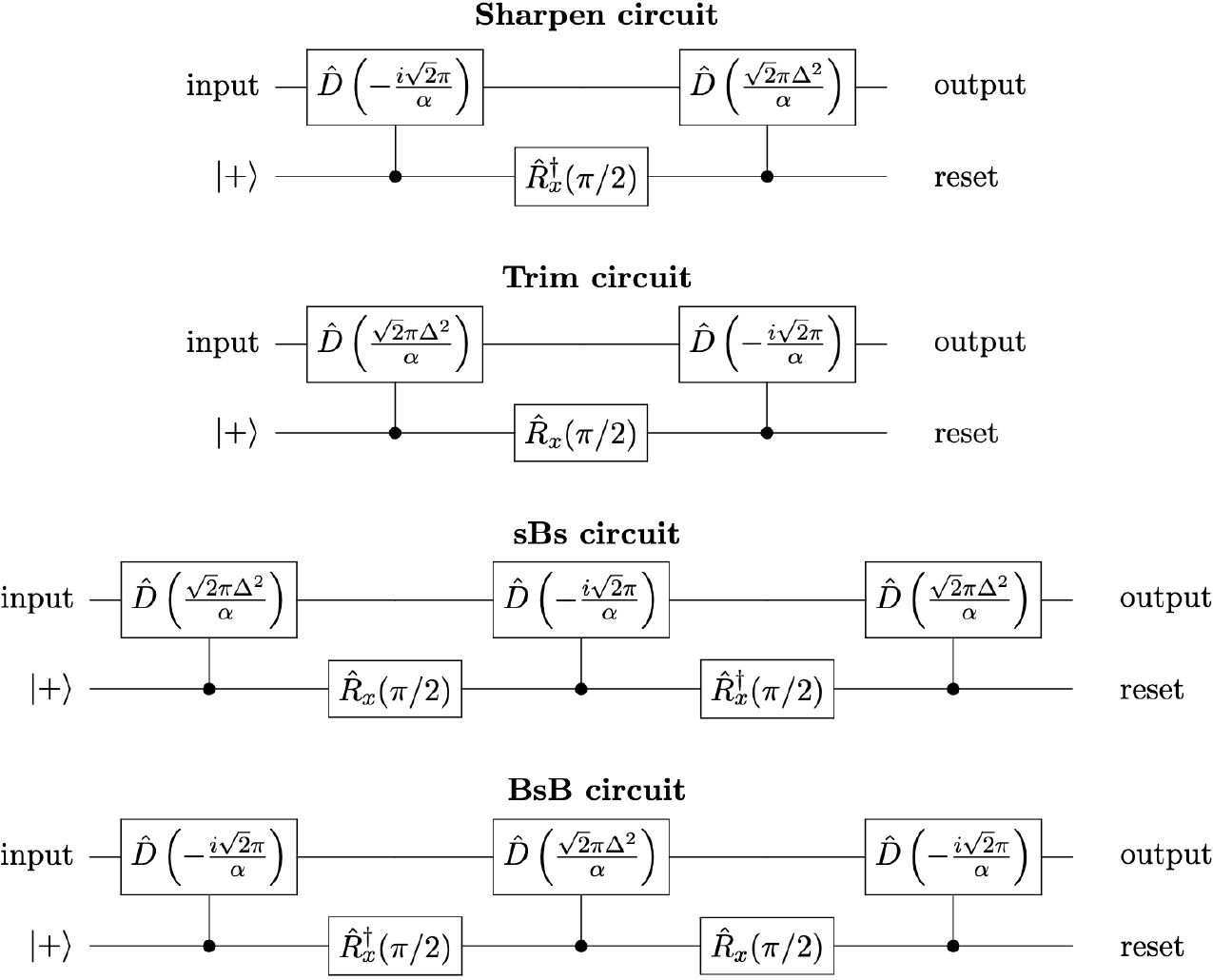}
    \caption{The circuits for stabilizing the squeezed cat code corresponding to the Sharpen, Trim, sBs, and BsB types of the Trotter decomposition. They consist of conditional displacement operations on the composite system with $X$-rotations on the ancillary qubit between them.}
    \label{fig:ST circuit}
\end{figure}

\textit{\textbf{Hardware-efficient autonomous error correction}}---
Here, we introduce the concrete autonomous QEC procedure for the dissipator $\hat{d}_{\Delta}$.
We construct a circuit that approximately realizes the dissipative process described by the Gorini–Kossakowski–Sudarshan–Lindblad (GKSL) equation $\frac{d}{dt} \h\rho= \gamma\calD[\hat{d}_{\Delta}](\h\rho) \coloneq \frac{\gamma}{2} (2 \hat{d}_{\Delta} \h\rho \hat{d}_{\Delta}^\dagger -\hat{d}_{\Delta}^\dagger \hat{d}_{\Delta} \h\rho - \h\rho \hat{d}_{\Delta}^\dagger \hat{d}_{\Delta})$ using an interaction with an ancillary qubit via conditional operations in a similar manner to that employed in Ref.~\cite{Royer2020} (see SM~\cite{Note1} for the detailed derivation).
The ``sharpen-trim'' (ST), ``small-Big-small'' (sBs), and ``Big-small-Big'' (BsB) unitary operators for the autonomous QEC are given as
\e{
U^{(\rm ST)} &=
\begin{cases}
    \ex^{-i\frac{\pi\Delta^2}{\sqrt{2} \alpha} \h{p} \otimes \h{\sigma}_y}\ex^{-i \frac{\pi}{\sqrt{2} \alpha} \h{x} \otimes \h{\sigma}_x} \quad {\text{(Sharpen)}}\\
    \ex^{-i\frac{\pi}{\sqrt{2} \alpha} \h{x} \otimes\h{\sigma}_x }\ex^{-i \frac{\pi \Delta^2}{\sqrt{2} \alpha} \h{p} \otimes \h{\sigma}_y}\quad {\text{(Trim)}}
\end{cases},\\
U^{(\rm sBs)} &= \ex^{-i\frac{\pi\Delta^2}{2\sqrt{2} \alpha} \h{p} \otimes \h{\sigma}_y}\ex^{-i \frac{\pi}{\sqrt{2} \alpha} \h{x} \otimes \h{\sigma}_x}\ex^{-i\frac{\pi\Delta^2}{2\sqrt{2} \alpha} \h{p} \otimes \h{\sigma}_y},\\
U^{(\rm BsB)} &= \ex^{-i \frac{\pi}{\sqrt{2} \alpha} \h{x} \otimes \h{\sigma}_x}\ex^{-i\frac{\sqrt{2}\pi\Delta^2}{ \alpha} \h{p} \otimes \h{\sigma}_y}\ex^{-i \frac{\pi}{\sqrt{2} \alpha} \h{x} \otimes \h{\sigma}_x},
}
with the ancillary qubit initialized to $\ket 0$.
Equivalent circuits are shown in Fig.~\ref{fig:ST circuit}.
Here, the conditional displacement operation is defined by $C \h{D}(\beta) \coloneqq \expo{(\beta a^\dagger - \beta^* a) \h{\sigma}_z / 2 \sqrt{2}}$, and the ancilla rotation is defined by $\h{R}_x(\theta) \coloneqq \expo{-i \theta \h{\sigma}_x / 2}$.
The circuit is hardware-friendly, since it only requires the conditional displacement operation, which is a standard operation in superconducting circuit QED systems~\cite{Campagne-Ibarcq2020,Eickbusch2022,Sivak2023,Lachance-Quirion2024} and trapped ion systems~\cite{Haljan2005,Fluhmann2018}.

\begin{figure}[htbp]
    \centering
    \includegraphics[width=\linewidth]{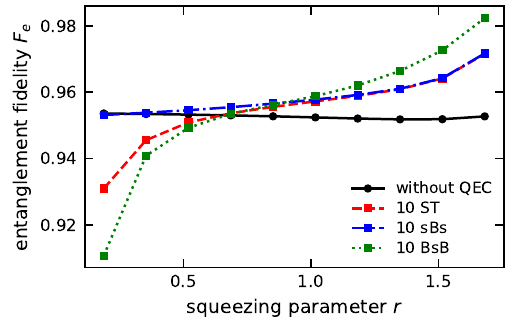}
    \caption{Entanglement fidelity $F_e$ against squeezing parameter $r$ after photon-loss noise with $\kappa t = 0.03$ followed by 10 applications of the ST (red), sBs (blue), BsB (green) protocol. Here, we take $\bar{n} = 10$.
    }
    \label{fig:fidelity_vs_sqlevel}
\end{figure}

To demonstrate the performance of the proposed QEC protocol, we perform a numerical simulation and demonstrate how the QEC protocol works against photon loss.
The bosonic mode undergoes the photon-loss process $\calE_{\rm loss}$ described by the GKSL equation $\od{}{\h\rho}{t} =  \kappa \calD[\h a](\h\rho)$ for the time interval satisfying $\kappa t = 0.01$ with $\kappa$ denoting the photon-loss rate of the bosonic mode.
Then, we apply the QEC circuit $m$ times, written as $\calE_{\rm QEC}^m$.
We evaluate the entanglement fidelity $F_e$ of the total process $\calE_{\rm QEC}^m\circ\calE_{\rm loss}$.
In Fig.~\ref{fig:fidelity_vs_sqlevel}, we plot the entanglement fidelity $F_e$ against the squeezing parameter $r$ while keeping the average number of photons $\bar{n}=\alpha^2 + \sinh^2 r$ fixed.
We see that the entanglement fidelity indeed recovers following application of the QEC circuit for sufficiently large values of the squeezing parameter $r$.
The BsB protocol involves time evolution that is twice as long as that of the other protocols. Therefore, for smaller values of r, the Trotter error becomes larger and the entanglement fidelity deteriorates. However, for larger values of r, more gauge excitations can be removed, leading to improved entanglement fidelity.
Since the sBs protocol corresponds to a second-order Trotter expansion and the Trotter error is sufficiently small, the entanglement fidelity does not deteriorate even for small r.

We have also numerically confirmed that our protocols are also effective for protecting code states from the dephasing error in SM~\cite{Note1}.

\begin{figure}[htbp]
    \centering
    \includegraphics[width=\linewidth]{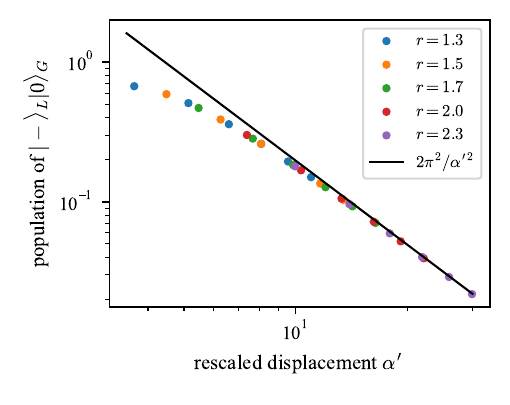}
    \caption{The population of $\ket -_L\ket 0_G$ after applying one cycle of ST to $\ket +_L\ket 1_G$, for different values of $\alpha\in\{1.0,1.4,1.8,2.2,2.6,3.0\}$ and $r\in\{1.3,1.5,1.7,2.0,2.3\}$. The subsystem decomposition analysis predicts it to be $2\pi^2 /\alpha'^2$, which is accurate for larger values of the rescaled displacement $\alpha'=\alpha e^r$.}
    \label{fig:TimeScale}
\end{figure}

\textit{\textbf{Efficiency and limitations}}---
We discuss efficiency and limitations of our protocol.
To estimate how rapidly our QEC circuit can remove excitations in the gauge mode, we note that the sharpen, trim, and sBs protocols each mimic the interaction with the environment over a time interval $\Gamma \delta t=\pi^2 /\alpha'^2$~\cite{Note1}, whereas the BsB protocol corresponds to a longer interval, $\Gamma \delta t=4\pi^2 /\alpha'^2$.
The interaction with the environmental qubit induces the dissipation $\calD[\h d_\Delta]$ on the bosonic mode for the same interval of $\Gamma t$, so that one cycle of the sharpen-trim circuit can eliminate $2\pi^2 /\alpha'^2$ excitations in the gauge mode.
To verify this, suppose that one sharpen-trim circuit is applied to a state $\ket{+}_L\ket 1_G$.
The ST circuit removes the excitation in the gauge mode with the probability $2\pi^2 /\alpha'^2$, accompanied by the phase-flip $Z_L$ in the logical space to generate the squeezed cat state $\ket{-}_L\ket 0_G$.
Figure~\ref{fig:TimeScale} shows the population of $\ket{-}_L\ket 0_G$, which is well-approximated by $2\pi^2 /\alpha'^2$ for large $\alpha'$.
For smaller values of $\alpha'$, the population deviates from the theoretical estimation because the time interval $\Gamma t$ is large, and the approximations such as the Trotterization of the interaction unitary operator and the replacement of the environmental bosonic mode with the qubit, become worse.

Let us consider a practical situation where the dominant noise is photon loss with a rate $\kappa$.
Then the photon loss induces excitations in gauge mode with the rate $\kappa \sinh^2 r$.
To remove them, the rate of applying the ST protocol should be greater than
\e{
\frac{\kappa \sinh^2 r}{2\pi^2 /\alpha'^2} \simeq \frac{\alpha'^2  e^{2r}}{8\pi^2}\kappa.
}

To discuss the limitations of our protocol, we note that in deriving the approximate expression in Eq.~\eqref{eq:dissipator} for the dissipator, we have utilized the fact that the action of the modular position operator on the gauge mode can be regarded as equivalent to that of the position operator for sufficiently large $\alpha'$.
Since the modular operator and the position operator differ only in their action outside the interval $[-\alpha'/\sqrt{2}, \alpha'/\sqrt{2}]$, this identification is valid only for states whose wavefunctions vanish outside this region.
The wavefunction of the Fock state $\braket{\tilde x|\tilde n}$ is mainly supported on the classically allowed region, i.e., $\frac{1}{2}\tilde x^2\le \tilde n + \frac{1}{2}\Leftrightarrow |\tilde x|\le \sqrt{2\tilde n + 1}$~\cite{Note1}. Indeed, the probability of detecting the position outside the classically allowed region vanishes asymptotically as $\sim \tilde n^{-1/3}$~\cite{Jadczyk2015, Paris2015}.
Therefore, the modular position in the gauge mode can be approximated by the position operator only for states spanned by the Fock states $\ket{\tilde n}$ in the gauge space with $\sqrt{2\tilde n +1} \lesssim \alpha'/\sqrt{2}$, or simply $\tilde n \lesssim \alpha'^2/4$.
This consideration explains why our QEC circuits can generate the SC state from the cat state, but not from the vacuum state (see SM~\cite{Note1} for details).

\textit{\textbf{Logical operations and Readout}}--- 
We can also exploit the translational symmetry and our QEC protocol to perform logical operations.
Indeed, we can implement the following:
$\mathcal P_{\ket +}$, the preparation of the $\ket{{\rm sq}_{\alpha,r}^+}$ state;
$X$, the logical Pauli $X$ operation;
$Z(\theta)$, the logical $Z$-rotation;
$ZZ(\theta)$, the logical $ZZ$-rotation;
and $\calM_Z$, the logical $Z$ readout.
We note that these operations constitute a universal set for quantum computation~\cite{Guillaud2019,Xu2022,Yuan2022,Puri2020,Xu2023}.
The implementation details for $\mathcal P_{\ket +}$ and $\calM_Z$ are given in End Matter, and details on the others are in SM~\cite{Note1}.


\textit{\textbf{Discussion and Conclusion}}---
In this work, we introduced a practical QEC protocol, logical operations, and measurement strategies for SC codes. We leveraged the unexplored non-unitary stabilizers from translational symmetries of SC codes for dissipative QEC.
Although the circuit for the QEC protocol is similar to the one realized in~\cite{Pan2023}, our protocol is autonomous and hence requires no feedback or conditioning.
It can also be used to mitigate errors during logical operations. In addition, we proposed an efficient measurement method for non-orthogonal basis measurement, which comes from the non-unitary stabilizer structure of SC codes. Finally we analytically and numerically verified that our protocol works in a hardware-efficient manner. 

Recent work has analyzed fault-tolerant performance obtained by concatenating a SC qubit with an outer code, including explicit surface-code constructions and logical-error estimates.
While the present work focuses on a hardware-efficient realization of autonomous correction for the SC manifold, the effective role of the bosonic layer is closely related: it suppresses loss-induced errors and enhances the resulting noise bias seen by the outer code. We therefore expect that, once mapped to a comparable effective noise model at the logical-qubit level, the concatenated logical-error performance will be broadly similar to that found in Ref.~\cite{Xu2023}. A full threshold study for concatenated architectures using the present QEC protocol is left for future work.

We have some possible future directions for this research. First, as an error suppression method using translational symmetries in bosonic codes, the projective squeezing method~\cite{Endo2024} has been proposed. While the projective squeezing method exploits post-selection for error suppression, the relationship with our proposed QEC is worth investigating---e.g., whether the subsystem representation can consistently describe the post-selection onto the code subspace.
Second, although the autonomous QEC had previously been proposed for GKP codes~\cite{Royer2020}, the behavior of the dissipative QEC on the code space has not yet been fully revealed. As in our research, by introducing an appropriate subsystem representation to the GKP dissipative QEC analysis, the functionality for reducing logical errors could be unraveled. 
Finally, although we choose the squeezed cat states as code words, it may be possible to use $\ket{\Psi_{\pm,n}}$ for $n\geq 1$ because the photon loss deterministically moves the quantum state outside the code subspace. Investigating the QEC capability of these quantum states could lead to a better construction and understanding of bosonic QEC with translational symmetries. 

\textit{\textbf{Acknowledgements}}---
TS thanks Ryuta Sasaki for a fruitful discussion on improved logical readout, and Ivan Rojkov for valuable comments on the earlier version of the manuscript.
GM acknowledges useful discussions with Dr. Isaac Chuang at the Massachusetts Institute of Technology.
The authors acknowledge the open-source software package QuTiP (Quantum Toolbox in Python)~\cite{Lambert2024} for numerical simulations performed in this work.
This work was supported by JST, CREST (Grant Nos. JPMJCR1771 and JPMJCR23I4), MEXT Q-LEAP (Grant Nos. JPMXS0120319794 and JPMXS0118068682), and JST Moonshot R\&D (Grant No. JPMJMS2061).

\makeatletter
\def\bibsection{}  
\makeatother

\bibliography{references.bib}

\begin{thebibliography}{43}%
\makeatletter
\providecommand \@ifxundefined [1]{%
 \@ifx{#1\undefined}
}%
\providecommand \@ifnum [1]{%
 \ifnum #1\expandafter \@firstoftwo
 \else \expandafter \@secondoftwo
 \fi
}%
\providecommand \@ifx [1]{%
 \ifx #1\expandafter \@firstoftwo
 \else \expandafter \@secondoftwo
 \fi
}%
\providecommand \natexlab [1]{#1}%
\providecommand \enquote  [1]{``#1''}%
\providecommand \bibnamefont  [1]{#1}%
\providecommand \bibfnamefont [1]{#1}%
\providecommand \citenamefont [1]{#1}%
\providecommand \href@noop [0]{\@secondoftwo}%
\providecommand \href [0]{\begingroup \@sanitize@url \@href}%
\providecommand \@href[1]{\@@startlink{#1}\@@href}%
\providecommand \@@href[1]{\endgroup#1\@@endlink}%
\providecommand \@sanitize@url [0]{\catcode `\\12\catcode `\$12\catcode `\&12\catcode `\#12\catcode `\^12\catcode `\_12\catcode `\%12\relax}%
\providecommand \@@startlink[1]{}%
\providecommand \@@endlink[0]{}%
\providecommand \url  [0]{\begingroup\@sanitize@url \@url }%
\providecommand \@url [1]{\endgroup\@href {#1}{\urlprefix }}%
\providecommand \urlprefix  [0]{URL }%
\providecommand \Eprint [0]{\href }%
\providecommand \doibase [0]{https://doi.org/}%
\providecommand \selectlanguage [0]{\@gobble}%
\providecommand \bibinfo  [0]{\@secondoftwo}%
\providecommand \bibfield  [0]{\@secondoftwo}%
\providecommand \translation [1]{[#1]}%
\providecommand \BibitemOpen [0]{}%
\providecommand \bibitemStop [0]{}%
\providecommand \bibitemNoStop [0]{.\EOS\space}%
\providecommand \EOS [0]{\spacefactor3000\relax}%
\providecommand \BibitemShut  [1]{\csname bibitem#1\endcsname}%
\let\auto@bib@innerbib\@empty
\bibitem [{\citenamefont {Shor}(1995)}]{Shor1995}%
  \BibitemOpen
  \bibfield  {author} {\bibinfo {author} {\bibfnamefont {P.~W.}\ \bibnamefont {Shor}},\ }\bibfield  {title} {\bibinfo {title} {{Scheme for reducing decoherence in quantum computer memory}},\ }\href@noop {} {\bibfield  {journal} {\bibinfo  {journal} {Physical Review A}\ }\textbf {\bibinfo {volume} {52}},\ \bibinfo {pages} {2493(R)} (\bibinfo {year} {1995})}\BibitemShut {NoStop}%
\bibitem [{\citenamefont {Bennett}\ \emph {et~al.}(1996)\citenamefont {Bennett}, \citenamefont {Divincenzo}, \citenamefont {Smolin},\ and\ \citenamefont {Wootters}}]{Bennett1996a}%
  \BibitemOpen
  \bibfield  {author} {\bibinfo {author} {\bibfnamefont {C.~H.}\ \bibnamefont {Bennett}}, \bibinfo {author} {\bibfnamefont {D.~P.}\ \bibnamefont {Divincenzo}}, \bibinfo {author} {\bibfnamefont {J.~A.}\ \bibnamefont {Smolin}},\ and\ \bibinfo {author} {\bibfnamefont {W.~K.}\ \bibnamefont {Wootters}},\ }\bibfield  {title} {\bibinfo {title} {{Mixed-state entanglement and quantum error correction}},\ }\href {https://doi.org/https://doi.org/10.1103/PhysRevA.54.3824} {\bibfield  {journal} {\bibinfo  {journal} {Physical Review A}\ }\textbf {\bibinfo {volume} {54}},\ \bibinfo {pages} {3824} (\bibinfo {year} {1996})}\BibitemShut {NoStop}%
\bibitem [{\citenamefont {Laflamme}\ \emph {et~al.}(1996)\citenamefont {Laflamme}, \citenamefont {Miquel}, \citenamefont {Paz},\ and\ \citenamefont {Zurek}}]{Laflamme1996}%
  \BibitemOpen
  \bibfield  {author} {\bibinfo {author} {\bibfnamefont {R.}~\bibnamefont {Laflamme}}, \bibinfo {author} {\bibfnamefont {C.}~\bibnamefont {Miquel}}, \bibinfo {author} {\bibfnamefont {J.~P.}\ \bibnamefont {Paz}},\ and\ \bibinfo {author} {\bibfnamefont {W.~H.}\ \bibnamefont {Zurek}},\ }\bibfield  {title} {\bibinfo {title} {{Perfect Quantum Error Correcting Code}},\ }\href@noop {} {\bibfield  {journal} {\bibinfo  {journal} {Physical Review Letters}\ }\textbf {\bibinfo {volume} {77}},\ \bibinfo {pages} {198} (\bibinfo {year} {1996})}\BibitemShut {NoStop}%
\bibitem [{\citenamefont {Knill}\ and\ \citenamefont {Laflamme}(1996)}]{Knill1996}%
  \BibitemOpen
  \bibfield  {author} {\bibinfo {author} {\bibfnamefont {E.}~\bibnamefont {Knill}}\ and\ \bibinfo {author} {\bibfnamefont {R.}~\bibnamefont {Laflamme}},\ }\bibfield  {title} {\bibinfo {title} {{Concatenated Quantum Codes}},\ }\href@noop {} {\bibfield  {journal} {\bibinfo  {journal} {arXiv}\ ,\ \bibinfo {pages} {quant}} (\bibinfo {year} {1996})}\BibitemShut {NoStop}%
\bibitem [{\citenamefont {{Michael A. Nielsen}}\ and\ \citenamefont {{Isaac L. Chuang}}(2000)}]{Nielsen2000}%
  \BibitemOpen
  \bibfield  {author} {\bibinfo {author} {\bibnamefont {{Michael A. Nielsen}}}\ and\ \bibinfo {author} {\bibnamefont {{Isaac L. Chuang}}},\ }\href@noop {} {\emph {\bibinfo {title} {{Quantum Computation and Quantum Information}}}}\ (\bibinfo  {publisher} {Cambridge University Press},\ \bibinfo {year} {2000})\BibitemShut {NoStop}%
\bibitem [{\citenamefont {Cochrane}\ \emph {et~al.}(1999)\citenamefont {Cochrane}, \citenamefont {Milburn},\ and\ \citenamefont {Munro}}]{Cochrane1999}%
  \BibitemOpen
  \bibfield  {author} {\bibinfo {author} {\bibfnamefont {P.~T.}\ \bibnamefont {Cochrane}}, \bibinfo {author} {\bibfnamefont {G.~J.}\ \bibnamefont {Milburn}},\ and\ \bibinfo {author} {\bibfnamefont {W.~J.}\ \bibnamefont {Munro}},\ }\bibfield  {title} {\bibinfo {title} {{Macroscopically distinct quantum-superposition states as a bosonic code for amplitude damping}},\ }\href@noop {} {\bibfield  {journal} {\bibinfo  {journal} {Physical Review A}\ }\textbf {\bibinfo {volume} {59}},\ \bibinfo {pages} {2631} (\bibinfo {year} {1999})}\BibitemShut {NoStop}%
\bibitem [{\citenamefont {Gottesman}\ \emph {et~al.}(2001)\citenamefont {Gottesman}, \citenamefont {Kitaev},\ and\ \citenamefont {Preskill}}]{Gottesman2001}%
  \BibitemOpen
  \bibfield  {author} {\bibinfo {author} {\bibfnamefont {D.}~\bibnamefont {Gottesman}}, \bibinfo {author} {\bibfnamefont {A.}~\bibnamefont {Kitaev}},\ and\ \bibinfo {author} {\bibfnamefont {J.}~\bibnamefont {Preskill}},\ }\bibfield  {title} {\bibinfo {title} {{Encoding a qubit in an oscillator}},\ }\href {https://doi.org/10.1103/PhysRevA.64.012310} {\bibfield  {journal} {\bibinfo  {journal} {Physical Review A}\ }\textbf {\bibinfo {volume} {64}},\ \bibinfo {pages} {123101} (\bibinfo {year} {2001})}\BibitemShut {NoStop}%
\bibitem [{\citenamefont {Michael}\ \emph {et~al.}(2016)\citenamefont {Michael}, \citenamefont {Silveri}, \citenamefont {Brierley}, \citenamefont {Albert}, \citenamefont {Salmilehto}, \citenamefont {Jiang},\ and\ \citenamefont {Girvin}}]{Michael2016}%
  \BibitemOpen
  \bibfield  {author} {\bibinfo {author} {\bibfnamefont {M.~H.}\ \bibnamefont {Michael}}, \bibinfo {author} {\bibfnamefont {M.}~\bibnamefont {Silveri}}, \bibinfo {author} {\bibfnamefont {R.~T.}\ \bibnamefont {Brierley}}, \bibinfo {author} {\bibfnamefont {V.~V.}\ \bibnamefont {Albert}}, \bibinfo {author} {\bibfnamefont {J.}~\bibnamefont {Salmilehto}}, \bibinfo {author} {\bibfnamefont {L.}~\bibnamefont {Jiang}},\ and\ \bibinfo {author} {\bibfnamefont {S.~M.}\ \bibnamefont {Girvin}},\ }\bibfield  {title} {\bibinfo {title} {{New class of quantum error-correcting codes for a bosonic mode}},\ }\href {https://doi.org/10.1103/PhysRevX.6.031006} {\bibfield  {journal} {\bibinfo  {journal} {Physical Review X}\ }\textbf {\bibinfo {volume} {6}},\ \bibinfo {pages} {031006} (\bibinfo {year} {2016})}\BibitemShut {NoStop}%
\bibitem [{\citenamefont {Grimsmo}\ \emph {et~al.}(2020)\citenamefont {Grimsmo}, \citenamefont {Combes},\ and\ \citenamefont {Baragiola}}]{Grimsmo2020}%
  \BibitemOpen
  \bibfield  {author} {\bibinfo {author} {\bibfnamefont {A.~L.}\ \bibnamefont {Grimsmo}}, \bibinfo {author} {\bibfnamefont {J.}~\bibnamefont {Combes}},\ and\ \bibinfo {author} {\bibfnamefont {B.~Q.}\ \bibnamefont {Baragiola}},\ }\bibfield  {title} {\bibinfo {title} {{Quantum Computing with Rotation-Symmetric Bosonic Codes}},\ }\href {https://doi.org/10.1103/PhysRevX.10.011058} {\bibfield  {journal} {\bibinfo  {journal} {Physical Review X}\ }\textbf {\bibinfo {volume} {10}},\ \bibinfo {pages} {011058} (\bibinfo {year} {2020})}\BibitemShut {NoStop}%
\bibitem [{\citenamefont {Walshe}\ \emph {et~al.}(2020)\citenamefont {Walshe}, \citenamefont {Baragiola}, \citenamefont {Alexander},\ and\ \citenamefont {Menicucci}}]{Walshe2020}%
  \BibitemOpen
  \bibfield  {author} {\bibinfo {author} {\bibfnamefont {B.~W.}\ \bibnamefont {Walshe}}, \bibinfo {author} {\bibfnamefont {B.~Q.}\ \bibnamefont {Baragiola}}, \bibinfo {author} {\bibfnamefont {R.~N.}\ \bibnamefont {Alexander}},\ and\ \bibinfo {author} {\bibfnamefont {N.~C.}\ \bibnamefont {Menicucci}},\ }\bibfield  {title} {\bibinfo {title} {{Continuous-variable gate teleportation and bosonic-code error correction}},\ }\href {https://doi.org/10.1103/PhysRevA.102.062411} {\bibfield  {journal} {\bibinfo  {journal} {Physical Review A}\ }\textbf {\bibinfo {volume} {102}},\ \bibinfo {pages} {062411} (\bibinfo {year} {2020})}\BibitemShut {NoStop}%
\bibitem [{\citenamefont {Noh}\ \emph {et~al.}(2020)\citenamefont {Noh}, \citenamefont {Girvin},\ and\ \citenamefont {Jiang}}]{Noh2020}%
  \BibitemOpen
  \bibfield  {author} {\bibinfo {author} {\bibfnamefont {K.}~\bibnamefont {Noh}}, \bibinfo {author} {\bibfnamefont {S.~M.}\ \bibnamefont {Girvin}},\ and\ \bibinfo {author} {\bibfnamefont {L.}~\bibnamefont {Jiang}},\ }\bibfield  {title} {\bibinfo {title} {{Encoding an Oscillator into Many Oscillators}},\ }\href {https://doi.org/10.1103/PhysRevLett.125.080503} {\bibfield  {journal} {\bibinfo  {journal} {Physical Review Letters}\ }\textbf {\bibinfo {volume} {125}},\ \bibinfo {pages} {080503} (\bibinfo {year} {2020})}\BibitemShut {NoStop}%
\bibitem [{\citenamefont {Royer}\ \emph {et~al.}(2022)\citenamefont {Royer}, \citenamefont {Singh},\ and\ \citenamefont {Girvin}}]{Royer2022}%
  \BibitemOpen
  \bibfield  {author} {\bibinfo {author} {\bibfnamefont {B.}~\bibnamefont {Royer}}, \bibinfo {author} {\bibfnamefont {S.}~\bibnamefont {Singh}},\ and\ \bibinfo {author} {\bibfnamefont {S.~M.}\ \bibnamefont {Girvin}},\ }\bibfield  {title} {\bibinfo {title} {{Encoding Qubits in Multimode Grid States}},\ }\href {https://doi.org/10.1103/PRXQuantum.3.010335} {\bibfield  {journal} {\bibinfo  {journal} {PRX Quantum}\ }\textbf {\bibinfo {volume} {3}},\ \bibinfo {pages} {010335} (\bibinfo {year} {2022})}\BibitemShut {NoStop}%
\bibitem [{\citenamefont {Mirrahimi}\ \emph {et~al.}(2014)\citenamefont {Mirrahimi}, \citenamefont {Leghtas}, \citenamefont {Albert}, \citenamefont {Touzard}, \citenamefont {Schoelkopf}, \citenamefont {Jiang},\ and\ \citenamefont {Devoret}}]{Mirrahimi2014}%
  \BibitemOpen
  \bibfield  {author} {\bibinfo {author} {\bibfnamefont {M.}~\bibnamefont {Mirrahimi}}, \bibinfo {author} {\bibfnamefont {Z.}~\bibnamefont {Leghtas}}, \bibinfo {author} {\bibfnamefont {V.~V.}\ \bibnamefont {Albert}}, \bibinfo {author} {\bibfnamefont {S.}~\bibnamefont {Touzard}}, \bibinfo {author} {\bibfnamefont {R.~J.}\ \bibnamefont {Schoelkopf}}, \bibinfo {author} {\bibfnamefont {L.}~\bibnamefont {Jiang}},\ and\ \bibinfo {author} {\bibfnamefont {M.~H.}\ \bibnamefont {Devoret}},\ }\bibfield  {title} {\bibinfo {title} {{Dynamically protected cat-qubits: A new paradigm for universal quantum computation}},\ }\href {https://doi.org/10.1088/1367-2630/16/4/045014} {\bibfield  {journal} {\bibinfo  {journal} {New Journal of Physics}\ }\textbf {\bibinfo {volume} {16}},\ \bibinfo {pages} {045014} (\bibinfo {year} {2014})}\BibitemShut {NoStop}%
\bibitem [{\citenamefont {Royer}\ \emph {et~al.}(2020)\citenamefont {Royer}, \citenamefont {Singh},\ and\ \citenamefont {Girvin}}]{Royer2020}%
  \BibitemOpen
  \bibfield  {author} {\bibinfo {author} {\bibfnamefont {B.}~\bibnamefont {Royer}}, \bibinfo {author} {\bibfnamefont {S.}~\bibnamefont {Singh}},\ and\ \bibinfo {author} {\bibfnamefont {S.~M.}\ \bibnamefont {Girvin}},\ }\bibfield  {title} {\bibinfo {title} {{Stabilization of Finite-Energy Gottesman-Kitaev-Preskill States}},\ }\href {https://doi.org/10.1103/PhysRevLett.125.260509} {\bibfield  {journal} {\bibinfo  {journal} {Physical Review Letters}\ }\textbf {\bibinfo {volume} {125}},\ \bibinfo {pages} {260509} (\bibinfo {year} {2020})}\BibitemShut {NoStop}%
\bibitem [{\citenamefont {Hillmann}\ and\ \citenamefont {Quijandr{\'{i}}a}(2023)}]{Hillmann2023}%
  \BibitemOpen
  \bibfield  {author} {\bibinfo {author} {\bibfnamefont {T.}~\bibnamefont {Hillmann}}\ and\ \bibinfo {author} {\bibfnamefont {F.}~\bibnamefont {Quijandr{\'{i}}a}},\ }\bibfield  {title} {\bibinfo {title} {{Quantum error correction with dissipatively stabilized squeezed-cat qubits}},\ }\href {https://doi.org/10.1103/PhysRevA.107.032423} {\bibfield  {journal} {\bibinfo  {journal} {Physical Review A}\ }\textbf {\bibinfo {volume} {107}},\ \bibinfo {pages} {032423} (\bibinfo {year} {2023})}\BibitemShut {NoStop}%
\bibitem [{\citenamefont {Schlegel}\ \emph {et~al.}(2022)\citenamefont {Schlegel}, \citenamefont {Minganti},\ and\ \citenamefont {Savona}}]{Schlegel2022}%
  \BibitemOpen
  \bibfield  {author} {\bibinfo {author} {\bibfnamefont {D.~S.}\ \bibnamefont {Schlegel}}, \bibinfo {author} {\bibfnamefont {F.}~\bibnamefont {Minganti}},\ and\ \bibinfo {author} {\bibfnamefont {V.}~\bibnamefont {Savona}},\ }\bibfield  {title} {\bibinfo {title} {{Quantum error correction using squeezed Schr{\"{o}}dinger cat states}},\ }\href {https://doi.org/10.1103/PhysRevA.106.022431} {\bibfield  {journal} {\bibinfo  {journal} {Physical Review A}\ }\textbf {\bibinfo {volume} {106}},\ \bibinfo {pages} {022431} (\bibinfo {year} {2022})}\BibitemShut {NoStop}%
\bibitem [{\citenamefont {Xu}\ \emph {et~al.}(2023)\citenamefont {Xu}, \citenamefont {Zheng}, \citenamefont {Wang}, \citenamefont {Zoller}, \citenamefont {Clerk},\ and\ \citenamefont {Jiang}}]{Xu2023}%
  \BibitemOpen
  \bibfield  {author} {\bibinfo {author} {\bibfnamefont {Q.}~\bibnamefont {Xu}}, \bibinfo {author} {\bibfnamefont {G.}~\bibnamefont {Zheng}}, \bibinfo {author} {\bibfnamefont {Y.-X.}\ \bibnamefont {Wang}}, \bibinfo {author} {\bibfnamefont {P.}~\bibnamefont {Zoller}}, \bibinfo {author} {\bibfnamefont {A.~A.}\ \bibnamefont {Clerk}},\ and\ \bibinfo {author} {\bibfnamefont {L.}~\bibnamefont {Jiang}},\ }\bibfield  {title} {\bibinfo {title} {{Autonomous quantum error correction and fault-tolerant quantum computation with squeezed cat qubits}},\ }\href {https://doi.org/10.1038/s41534-023-00746-0} {\bibfield  {journal} {\bibinfo  {journal} {npj Quantum Information}\ }\textbf {\bibinfo {volume} {9}},\ \bibinfo {pages} {78} (\bibinfo {year} {2023})}\BibitemShut {NoStop}%
\bibitem [{\citenamefont {Putterman}\ \emph {et~al.}(2022)\citenamefont {Putterman}, \citenamefont {Iverson}, \citenamefont {Xu}, \citenamefont {Jiang}, \citenamefont {Painter}, \citenamefont {Brand{\~{a}}o},\ and\ \citenamefont {Noh}}]{Putterman2022}%
  \BibitemOpen
  \bibfield  {author} {\bibinfo {author} {\bibfnamefont {H.}~\bibnamefont {Putterman}}, \bibinfo {author} {\bibfnamefont {J.}~\bibnamefont {Iverson}}, \bibinfo {author} {\bibfnamefont {Q.}~\bibnamefont {Xu}}, \bibinfo {author} {\bibfnamefont {L.}~\bibnamefont {Jiang}}, \bibinfo {author} {\bibfnamefont {O.}~\bibnamefont {Painter}}, \bibinfo {author} {\bibfnamefont {F.~G.}\ \bibnamefont {Brand{\~{a}}o}},\ and\ \bibinfo {author} {\bibfnamefont {K.}~\bibnamefont {Noh}},\ }\bibfield  {title} {\bibinfo {title} {{Stabilizing a Bosonic Qubit Using Colored Dissipation}},\ }\href {https://doi.org/10.1103/PhysRevLett.128.110502} {\bibfield  {journal} {\bibinfo  {journal} {Physical Review Letters}\ }\textbf {\bibinfo {volume} {128}},\ \bibinfo {pages} {110502} (\bibinfo {year} {2022})}\BibitemShut {NoStop}%
\bibitem [{\citenamefont {Chamberland}\ \emph {et~al.}(2022)\citenamefont {Chamberland}, \citenamefont {Noh}, \citenamefont {Arrangoiz-Arriola}, \citenamefont {Campbell}, \citenamefont {Hann}, \citenamefont {Iverson}, \citenamefont {Putterman}, \citenamefont {Bohdanowicz}, \citenamefont {Flammia}, \citenamefont {Keller}, \citenamefont {Refael}, \citenamefont {Preskill}, \citenamefont {Jiang}, \citenamefont {Safavi-Naeini}, \citenamefont {Painter},\ and\ \citenamefont {Brand{\~{a}}o}}]{Chamberland2022}%
  \BibitemOpen
  \bibfield  {author} {\bibinfo {author} {\bibfnamefont {C.}~\bibnamefont {Chamberland}}, \bibinfo {author} {\bibfnamefont {K.}~\bibnamefont {Noh}}, \bibinfo {author} {\bibfnamefont {P.}~\bibnamefont {Arrangoiz-Arriola}}, \bibinfo {author} {\bibfnamefont {E.~T.}\ \bibnamefont {Campbell}}, \bibinfo {author} {\bibfnamefont {C.~T.}\ \bibnamefont {Hann}}, \bibinfo {author} {\bibfnamefont {J.}~\bibnamefont {Iverson}}, \bibinfo {author} {\bibfnamefont {H.}~\bibnamefont {Putterman}}, \bibinfo {author} {\bibfnamefont {T.~C.}\ \bibnamefont {Bohdanowicz}}, \bibinfo {author} {\bibfnamefont {S.~T.}\ \bibnamefont {Flammia}}, \bibinfo {author} {\bibfnamefont {A.}~\bibnamefont {Keller}}, \bibinfo {author} {\bibfnamefont {G.}~\bibnamefont {Refael}}, \bibinfo {author} {\bibfnamefont {J.}~\bibnamefont {Preskill}}, \bibinfo {author} {\bibfnamefont {L.}~\bibnamefont {Jiang}}, \bibinfo {author} {\bibfnamefont {A.~H.}\ \bibnamefont {Safavi-Naeini}}, \bibinfo {author} {\bibfnamefont {O.}~\bibnamefont {Painter}},\ and\ \bibinfo
  {author} {\bibfnamefont {F.~G.}\ \bibnamefont {Brand{\~{a}}o}},\ }\bibfield  {title} {\bibinfo {title} {{Building a Fault-Tolerant Quantum Computer Using Concatenated Cat Codes}},\ }\href {https://doi.org/10.1103/PRXQuantum.3.010329} {\bibfield  {journal} {\bibinfo  {journal} {PRX Quantum}\ }\textbf {\bibinfo {volume} {3}},\ \bibinfo {pages} {010329} (\bibinfo {year} {2022})}\BibitemShut {NoStop}%
\bibitem [{\citenamefont {Endo}\ \emph {et~al.}(2024)\citenamefont {Endo}, \citenamefont {Anai}, \citenamefont {Matsuzaki}, \citenamefont {Tokunaga},\ and\ \citenamefont {Suzuki}}]{Endo2024}%
  \BibitemOpen
  \bibfield  {author} {\bibinfo {author} {\bibfnamefont {S.}~\bibnamefont {Endo}}, \bibinfo {author} {\bibfnamefont {K.}~\bibnamefont {Anai}}, \bibinfo {author} {\bibfnamefont {Y.}~\bibnamefont {Matsuzaki}}, \bibinfo {author} {\bibfnamefont {Y.}~\bibnamefont {Tokunaga}},\ and\ \bibinfo {author} {\bibfnamefont {Y.}~\bibnamefont {Suzuki}},\ }\bibfield  {title} {\bibinfo {title} {{Projective squeezing for translation symmetric bosonic codes}},\ }\href {http://arxiv.org/abs/2403.14218} {\bibfield  {journal} {\bibinfo  {journal} {arXiv}\ ,\ \bibinfo {pages} {2403.14218}} (\bibinfo {year} {2024})}\BibitemShut {NoStop}%
\bibitem [{\citenamefont {Xu}\ \emph {et~al.}(2022)\citenamefont {Xu}, \citenamefont {Iverson}, \citenamefont {Brand{\~{a}}o},\ and\ \citenamefont {Jiang}}]{Xu2022}%
  \BibitemOpen
  \bibfield  {author} {\bibinfo {author} {\bibfnamefont {Q.}~\bibnamefont {Xu}}, \bibinfo {author} {\bibfnamefont {J.~K.}\ \bibnamefont {Iverson}}, \bibinfo {author} {\bibfnamefont {F.~G.}\ \bibnamefont {Brand{\~{a}}o}},\ and\ \bibinfo {author} {\bibfnamefont {L.}~\bibnamefont {Jiang}},\ }\bibfield  {title} {\bibinfo {title} {{Engineering fast bias-preserving gates on stabilized cat qubits}},\ }\href {https://doi.org/10.1103/PhysRevResearch.4.013082} {\bibfield  {journal} {\bibinfo  {journal} {Physical Review Research}\ }\textbf {\bibinfo {volume} {4}},\ \bibinfo {pages} {013082} (\bibinfo {year} {2022})}\BibitemShut {NoStop}%
\bibitem [{Note1()}]{Note1}%
  \BibitemOpen
  \bibinfo {note} {See Supplemental Material.}\BibitemShut {Stop}%
\bibitem [{Note2()}]{Note2}%
  \BibitemOpen
  \bibinfo {note} {The dissipator is chosen so that $\protect \hat {d}_\Delta \propto \log \protect \hat {T}_\Delta $ and it is normalized in the sense that it behaves like an annihilation operator if we neglect the modularity---that is, $\lim _{\alpha \to \infty }[\protect \hat {d}_\Delta , \protect \hat {d}_\Delta ^\dagger ] = \protect \hat {I}$.}\BibitemShut {Stop}%
\bibitem [{\citenamefont {Zak}(1967)}]{Zak1967}%
  \BibitemOpen
  \bibfield  {author} {\bibinfo {author} {\bibfnamefont {J.}~\bibnamefont {Zak}},\ }\bibfield  {title} {\bibinfo {title} {{Finite Translations in Solid-State Physics}},\ }\href@noop {} {\bibfield  {journal} {\bibinfo  {journal} {Physical Review Letters}\ }\textbf {\bibinfo {volume} {19}},\ \bibinfo {pages} {1385} (\bibinfo {year} {1967})}\BibitemShut {NoStop}%
\bibitem [{\citenamefont {Pantaleoni}\ \emph {et~al.}(2023)\citenamefont {Pantaleoni}, \citenamefont {Baragiola},\ and\ \citenamefont {Menicucci}}]{Pantaleoni2023}%
  \BibitemOpen
  \bibfield  {author} {\bibinfo {author} {\bibfnamefont {G.}~\bibnamefont {Pantaleoni}}, \bibinfo {author} {\bibfnamefont {B.~Q.}\ \bibnamefont {Baragiola}},\ and\ \bibinfo {author} {\bibfnamefont {N.~C.}\ \bibnamefont {Menicucci}},\ }\bibfield  {title} {\bibinfo {title} {{Zak transform as a framework for quantum computation with the Gottesman-Kitaev-Preskill code}},\ }\href {https://doi.org/10.1103/PhysRevA.107.062611} {\bibfield  {journal} {\bibinfo  {journal} {Physical Review A}\ }\textbf {\bibinfo {volume} {107}},\ \bibinfo {pages} {062611} (\bibinfo {year} {2023})}\BibitemShut {NoStop}%
\bibitem [{\citenamefont {Campagne-Ibarcq}\ \emph {et~al.}(2020)\citenamefont {Campagne-Ibarcq}, \citenamefont {Eickbusch}, \citenamefont {Touzard}, \citenamefont {Zalys-Geller}, \citenamefont {Frattini}, \citenamefont {Sivak}, \citenamefont {Reinhold}, \citenamefont {Puri}, \citenamefont {Shankar}, \citenamefont {Schoelkopf}, \citenamefont {Frunzio}, \citenamefont {Mirrahimi},\ and\ \citenamefont {Devoret}}]{Campagne-Ibarcq2020}%
  \BibitemOpen
  \bibfield  {author} {\bibinfo {author} {\bibfnamefont {P.}~\bibnamefont {Campagne-Ibarcq}}, \bibinfo {author} {\bibfnamefont {A.}~\bibnamefont {Eickbusch}}, \bibinfo {author} {\bibfnamefont {S.}~\bibnamefont {Touzard}}, \bibinfo {author} {\bibfnamefont {E.}~\bibnamefont {Zalys-Geller}}, \bibinfo {author} {\bibfnamefont {N.~E.}\ \bibnamefont {Frattini}}, \bibinfo {author} {\bibfnamefont {V.~V.}\ \bibnamefont {Sivak}}, \bibinfo {author} {\bibfnamefont {P.}~\bibnamefont {Reinhold}}, \bibinfo {author} {\bibfnamefont {S.}~\bibnamefont {Puri}}, \bibinfo {author} {\bibfnamefont {S.}~\bibnamefont {Shankar}}, \bibinfo {author} {\bibfnamefont {R.~J.}\ \bibnamefont {Schoelkopf}}, \bibinfo {author} {\bibfnamefont {L.}~\bibnamefont {Frunzio}}, \bibinfo {author} {\bibfnamefont {M.}~\bibnamefont {Mirrahimi}},\ and\ \bibinfo {author} {\bibfnamefont {M.~H.}\ \bibnamefont {Devoret}},\ }\bibfield  {title} {\bibinfo {title} {{Quantum error correction of a qubit encoded in grid states of an oscillator}},\ }\href
  {https://doi.org/10.1038/s41586-020-2603-3} {\bibfield  {journal} {\bibinfo  {journal} {Nature}\ }\textbf {\bibinfo {volume} {584}},\ \bibinfo {pages} {368} (\bibinfo {year} {2020})}\BibitemShut {NoStop}%
\bibitem [{\citenamefont {Eickbusch}\ \emph {et~al.}(2022)\citenamefont {Eickbusch}, \citenamefont {Sivak}, \citenamefont {Ding}, \citenamefont {Elder}, \citenamefont {Jha}, \citenamefont {Venkatraman}, \citenamefont {Royer}, \citenamefont {Girvin}, \citenamefont {Schoelkopf},\ and\ \citenamefont {Devoret}}]{Eickbusch2022}%
  \BibitemOpen
  \bibfield  {author} {\bibinfo {author} {\bibfnamefont {A.}~\bibnamefont {Eickbusch}}, \bibinfo {author} {\bibfnamefont {V.}~\bibnamefont {Sivak}}, \bibinfo {author} {\bibfnamefont {A.~Z.}\ \bibnamefont {Ding}}, \bibinfo {author} {\bibfnamefont {S.~S.}\ \bibnamefont {Elder}}, \bibinfo {author} {\bibfnamefont {S.~R.}\ \bibnamefont {Jha}}, \bibinfo {author} {\bibfnamefont {J.}~\bibnamefont {Venkatraman}}, \bibinfo {author} {\bibfnamefont {B.}~\bibnamefont {Royer}}, \bibinfo {author} {\bibfnamefont {S.~M.}\ \bibnamefont {Girvin}}, \bibinfo {author} {\bibfnamefont {R.~J.}\ \bibnamefont {Schoelkopf}},\ and\ \bibinfo {author} {\bibfnamefont {M.~H.}\ \bibnamefont {Devoret}},\ }\bibfield  {title} {\bibinfo {title} {{Fast universal control of an oscillator with weak dispersive coupling to a qubit}},\ }\href {https://doi.org/10.1038/s41567-022-01776-9} {\bibfield  {journal} {\bibinfo  {journal} {Nature Physics}\ }\textbf {\bibinfo {volume} {18}},\ \bibinfo {pages} {1464} (\bibinfo {year} {2022})}\BibitemShut {NoStop}%
\bibitem [{\citenamefont {Sivak}\ \emph {et~al.}(2023)\citenamefont {Sivak}, \citenamefont {Eickbusch}, \citenamefont {Royer}, \citenamefont {Singh}, \citenamefont {Tsioutsios}, \citenamefont {Ganjam}, \citenamefont {Miano}, \citenamefont {Brock}, \citenamefont {Ding}, \citenamefont {Frunzio}, \citenamefont {Girvin}, \citenamefont {Schoelkopf},\ and\ \citenamefont {Devoret}}]{Sivak2023}%
  \BibitemOpen
  \bibfield  {author} {\bibinfo {author} {\bibfnamefont {V.~V.}\ \bibnamefont {Sivak}}, \bibinfo {author} {\bibfnamefont {A.}~\bibnamefont {Eickbusch}}, \bibinfo {author} {\bibfnamefont {B.}~\bibnamefont {Royer}}, \bibinfo {author} {\bibfnamefont {S.}~\bibnamefont {Singh}}, \bibinfo {author} {\bibfnamefont {I.}~\bibnamefont {Tsioutsios}}, \bibinfo {author} {\bibfnamefont {S.}~\bibnamefont {Ganjam}}, \bibinfo {author} {\bibfnamefont {A.}~\bibnamefont {Miano}}, \bibinfo {author} {\bibfnamefont {B.~L.}\ \bibnamefont {Brock}}, \bibinfo {author} {\bibfnamefont {A.~Z.}\ \bibnamefont {Ding}}, \bibinfo {author} {\bibfnamefont {L.}~\bibnamefont {Frunzio}}, \bibinfo {author} {\bibfnamefont {S.~M.}\ \bibnamefont {Girvin}}, \bibinfo {author} {\bibfnamefont {R.~J.}\ \bibnamefont {Schoelkopf}},\ and\ \bibinfo {author} {\bibfnamefont {M.~H.}\ \bibnamefont {Devoret}},\ }\bibfield  {title} {\bibinfo {title} {{Real-time quantum error correction beyond break-even}},\ }\href {https://doi.org/10.1038/s41586-023-05782-6}
  {\bibfield  {journal} {\bibinfo  {journal} {Nature}\ }\textbf {\bibinfo {volume} {616}},\ \bibinfo {pages} {50} (\bibinfo {year} {2023})}\BibitemShut {NoStop}%
\bibitem [{\citenamefont {Lachance-Quirion}\ \emph {et~al.}(2024)\citenamefont {Lachance-Quirion}, \citenamefont {Lemonde}, \citenamefont {Simoneau}, \citenamefont {St-Jean}, \citenamefont {Lemieux}, \citenamefont {Turcotte}, \citenamefont {Wright}, \citenamefont {Lacroix}, \citenamefont {Fr{\'{e}}chette-Viens}, \citenamefont {Shillito}, \citenamefont {Hopfmueller}, \citenamefont {Tremblay}, \citenamefont {Frattini}, \citenamefont {Camirand~Lemyre},\ and\ \citenamefont {St-Jean}}]{Lachance-Quirion2024}%
  \BibitemOpen
  \bibfield  {author} {\bibinfo {author} {\bibfnamefont {D.}~\bibnamefont {Lachance-Quirion}}, \bibinfo {author} {\bibfnamefont {M.~A.}\ \bibnamefont {Lemonde}}, \bibinfo {author} {\bibfnamefont {J.~O.}\ \bibnamefont {Simoneau}}, \bibinfo {author} {\bibfnamefont {L.}~\bibnamefont {St-Jean}}, \bibinfo {author} {\bibfnamefont {P.}~\bibnamefont {Lemieux}}, \bibinfo {author} {\bibfnamefont {S.}~\bibnamefont {Turcotte}}, \bibinfo {author} {\bibfnamefont {W.}~\bibnamefont {Wright}}, \bibinfo {author} {\bibfnamefont {A.}~\bibnamefont {Lacroix}}, \bibinfo {author} {\bibfnamefont {J.}~\bibnamefont {Fr{\'{e}}chette-Viens}}, \bibinfo {author} {\bibfnamefont {R.}~\bibnamefont {Shillito}}, \bibinfo {author} {\bibfnamefont {F.}~\bibnamefont {Hopfmueller}}, \bibinfo {author} {\bibfnamefont {M.}~\bibnamefont {Tremblay}}, \bibinfo {author} {\bibfnamefont {N.~E.}\ \bibnamefont {Frattini}}, \bibinfo {author} {\bibfnamefont {J.}~\bibnamefont {Camirand~Lemyre}},\ and\ \bibinfo {author} {\bibfnamefont {P.}~\bibnamefont
  {St-Jean}},\ }\bibfield  {title} {\bibinfo {title} {{Autonomous Quantum Error Correction of Gottesman-Kitaev-Preskill States}},\ }\href {https://doi.org/10.1103/PhysRevLett.132.150607} {\bibfield  {journal} {\bibinfo  {journal} {Physical Review Letters}\ }\textbf {\bibinfo {volume} {132}},\ \bibinfo {pages} {150607} (\bibinfo {year} {2024})}\BibitemShut {NoStop}%
\bibitem [{\citenamefont {Haljan}\ \emph {et~al.}(2005)\citenamefont {Haljan}, \citenamefont {Brickman}, \citenamefont {Deslauriers}, \citenamefont {Lee},\ and\ \citenamefont {Monroe}}]{Haljan2005}%
  \BibitemOpen
  \bibfield  {author} {\bibinfo {author} {\bibfnamefont {P.~C.}\ \bibnamefont {Haljan}}, \bibinfo {author} {\bibfnamefont {K.~A.}\ \bibnamefont {Brickman}}, \bibinfo {author} {\bibfnamefont {L.}~\bibnamefont {Deslauriers}}, \bibinfo {author} {\bibfnamefont {P.~J.}\ \bibnamefont {Lee}},\ and\ \bibinfo {author} {\bibfnamefont {C.}~\bibnamefont {Monroe}},\ }\bibfield  {title} {\bibinfo {title} {{Spin-dependent forces on trapped ions for phase-stable quantum gates and entangled states of spin and motion}},\ }\href {https://doi.org/10.1103/PhysRevLett.94.153602} {\bibfield  {journal} {\bibinfo  {journal} {Physical Review Letters}\ }\textbf {\bibinfo {volume} {94}},\ \bibinfo {pages} {153602} (\bibinfo {year} {2005})}\BibitemShut {NoStop}%
\bibitem [{\citenamefont {Fl{\"{u}}hmann}\ \emph {et~al.}(2018)\citenamefont {Fl{\"{u}}hmann}, \citenamefont {Negnevitsky}, \citenamefont {Marinelli},\ and\ \citenamefont {Home}}]{Fluhmann2018}%
  \BibitemOpen
  \bibfield  {author} {\bibinfo {author} {\bibfnamefont {C.}~\bibnamefont {Fl{\"{u}}hmann}}, \bibinfo {author} {\bibfnamefont {V.}~\bibnamefont {Negnevitsky}}, \bibinfo {author} {\bibfnamefont {M.}~\bibnamefont {Marinelli}},\ and\ \bibinfo {author} {\bibfnamefont {J.~P.}\ \bibnamefont {Home}},\ }\bibfield  {title} {\bibinfo {title} {{Sequential Modular Position and Momentum Measurements of a Trapped Ion Mechanical Oscillator}},\ }\href {https://doi.org/10.1103/PhysRevX.8.021001} {\bibfield  {journal} {\bibinfo  {journal} {Physical Review X}\ }\textbf {\bibinfo {volume} {8}},\ \bibinfo {pages} {021001} (\bibinfo {year} {2018})}\BibitemShut {NoStop}%
\bibitem [{\citenamefont {Jadczyk}(2015)}]{Jadczyk2015}%
  \BibitemOpen
  \bibfield  {author} {\bibinfo {author} {\bibfnamefont {A.}~\bibnamefont {Jadczyk}},\ }\bibfield  {title} {\bibinfo {title} {{Asymptotic formula for quantum harmonic oscillator tunneling probabilities}},\ }\href {https://doi.org/10.1016/S0034-4877(15)30025-2} {\bibfield  {journal} {\bibinfo  {journal} {Reports on Mathematical Physics}\ }\textbf {\bibinfo {volume} {76}},\ \bibinfo {pages} {149} (\bibinfo {year} {2015})}\BibitemShut {NoStop}%
\bibitem [{\citenamefont {Paris}(2015)}]{Paris2015}%
  \BibitemOpen
  \bibfield  {author} {\bibinfo {author} {\bibfnamefont {R.~B.}\ \bibnamefont {Paris}},\ }\bibfield  {title} {\bibinfo {title} {{Asymptotic evaluation of an integral arising in quantum harmonic oscillator tunnelling probabilities}},\ }\href {http://arxiv.org/abs/1502.03382} {\bibfield  {journal} {\bibinfo  {journal} {arXiv}\ ,\ \bibinfo {pages} {1502.03382}} (\bibinfo {year} {2015})}\BibitemShut {NoStop}%
\bibitem [{\citenamefont {Guillaud}\ and\ \citenamefont {Mirrahimi}(2019)}]{Guillaud2019}%
  \BibitemOpen
  \bibfield  {author} {\bibinfo {author} {\bibfnamefont {J.}~\bibnamefont {Guillaud}}\ and\ \bibinfo {author} {\bibfnamefont {M.}~\bibnamefont {Mirrahimi}},\ }\bibfield  {title} {\bibinfo {title} {{Repetition Cat Qubits for Fault-Tolerant Quantum Computation}},\ }\href {https://doi.org/10.1103/PhysRevX.9.041053} {\bibfield  {journal} {\bibinfo  {journal} {Physical Review X}\ }\textbf {\bibinfo {volume} {9}},\ \bibinfo {pages} {041053} (\bibinfo {year} {2019})}\BibitemShut {NoStop}%
\bibitem [{\citenamefont {Yuan}\ \emph {et~al.}(2022)\citenamefont {Yuan}, \citenamefont {Xu},\ and\ \citenamefont {Jiang}}]{Yuan2022}%
  \BibitemOpen
  \bibfield  {author} {\bibinfo {author} {\bibfnamefont {M.}~\bibnamefont {Yuan}}, \bibinfo {author} {\bibfnamefont {Q.}~\bibnamefont {Xu}},\ and\ \bibinfo {author} {\bibfnamefont {L.}~\bibnamefont {Jiang}},\ }\bibfield  {title} {\bibinfo {title} {{Construction of bias-preserving operations for pair-cat codes}},\ }\href {https://doi.org/10.1103/PhysRevA.106.062422} {\bibfield  {journal} {\bibinfo  {journal} {Physical Review A}\ }\textbf {\bibinfo {volume} {106}},\ \bibinfo {pages} {062422} (\bibinfo {year} {2022})}\BibitemShut {NoStop}%
\bibitem [{\citenamefont {Puri}\ \emph {et~al.}(2020)\citenamefont {Puri}, \citenamefont {St-Jean}, \citenamefont {Gross}, \citenamefont {Grimm}, \citenamefont {Frattini}, \citenamefont {Iyer}, \citenamefont {Krishna}, \citenamefont {Touzard}, \citenamefont {Jiang}, \citenamefont {Blais}, \citenamefont {Flammia},\ and\ \citenamefont {Girvin}}]{Puri2020}%
  \BibitemOpen
  \bibfield  {author} {\bibinfo {author} {\bibfnamefont {S.}~\bibnamefont {Puri}}, \bibinfo {author} {\bibfnamefont {L.}~\bibnamefont {St-Jean}}, \bibinfo {author} {\bibfnamefont {J.~A.}\ \bibnamefont {Gross}}, \bibinfo {author} {\bibfnamefont {A.}~\bibnamefont {Grimm}}, \bibinfo {author} {\bibfnamefont {N.~E.}\ \bibnamefont {Frattini}}, \bibinfo {author} {\bibfnamefont {P.~S.}\ \bibnamefont {Iyer}}, \bibinfo {author} {\bibfnamefont {A.}~\bibnamefont {Krishna}}, \bibinfo {author} {\bibfnamefont {S.}~\bibnamefont {Touzard}}, \bibinfo {author} {\bibfnamefont {L.}~\bibnamefont {Jiang}}, \bibinfo {author} {\bibfnamefont {A.}~\bibnamefont {Blais}}, \bibinfo {author} {\bibfnamefont {S.~T.}\ \bibnamefont {Flammia}},\ and\ \bibinfo {author} {\bibfnamefont {S.~M.}\ \bibnamefont {Girvin}},\ }\bibfield  {title} {\bibinfo {title} {{Bias-preserving gates with stabilized cat qubits}},\ }\href {https://www.science.org} {\bibfield  {journal} {\bibinfo  {journal} {Science Advances}\ }\textbf {\bibinfo {volume} {6}},\ \bibinfo
  {pages} {5901} (\bibinfo {year} {2020})}\BibitemShut {NoStop}%
\bibitem [{\citenamefont {Pan}\ \emph {et~al.}(2023)\citenamefont {Pan}, \citenamefont {Schwinger}, \citenamefont {Huang}, \citenamefont {Song}, \citenamefont {Chua}, \citenamefont {Hanamura}, \citenamefont {Joshi}, \citenamefont {Valadares}, \citenamefont {Filip},\ and\ \citenamefont {Gao}}]{Pan2023}%
  \BibitemOpen
  \bibfield  {author} {\bibinfo {author} {\bibfnamefont {X.}~\bibnamefont {Pan}}, \bibinfo {author} {\bibfnamefont {J.}~\bibnamefont {Schwinger}}, \bibinfo {author} {\bibfnamefont {N.~N.}\ \bibnamefont {Huang}}, \bibinfo {author} {\bibfnamefont {P.}~\bibnamefont {Song}}, \bibinfo {author} {\bibfnamefont {W.}~\bibnamefont {Chua}}, \bibinfo {author} {\bibfnamefont {F.}~\bibnamefont {Hanamura}}, \bibinfo {author} {\bibfnamefont {A.}~\bibnamefont {Joshi}}, \bibinfo {author} {\bibfnamefont {F.}~\bibnamefont {Valadares}}, \bibinfo {author} {\bibfnamefont {R.}~\bibnamefont {Filip}},\ and\ \bibinfo {author} {\bibfnamefont {Y.~Y.}\ \bibnamefont {Gao}},\ }\bibfield  {title} {\bibinfo {title} {{Protecting the Quantum Interference of Cat States by Phase-Space Compression}},\ }\href {https://doi.org/10.1103/PhysRevX.13.021004} {\bibfield  {journal} {\bibinfo  {journal} {Physical Review X}\ }\textbf {\bibinfo {volume} {13}},\ \bibinfo {pages} {021004} (\bibinfo {year} {2023})}\BibitemShut {NoStop}%
\bibitem [{\citenamefont {Lambert}\ \emph {et~al.}(2024)\citenamefont {Lambert}, \citenamefont {Gigu{\`{e}}re}, \citenamefont {Menczel}, \citenamefont {Li}, \citenamefont {Hopf}, \citenamefont {Su{\'{a}}rez}, \citenamefont {Gali}, \citenamefont {Lishman}, \citenamefont {Gadhvi}, \citenamefont {Agarwal}, \citenamefont {Galicia}, \citenamefont {Shammah}, \citenamefont {Nation}, \citenamefont {Johansson}, \citenamefont {Ahmed}, \citenamefont {Cross}, \citenamefont {Pitchford},\ and\ \citenamefont {Nori}}]{Lambert2024}%
  \BibitemOpen
  \bibfield  {author} {\bibinfo {author} {\bibfnamefont {N.}~\bibnamefont {Lambert}}, \bibinfo {author} {\bibfnamefont {E.}~\bibnamefont {Gigu{\`{e}}re}}, \bibinfo {author} {\bibfnamefont {P.}~\bibnamefont {Menczel}}, \bibinfo {author} {\bibfnamefont {B.}~\bibnamefont {Li}}, \bibinfo {author} {\bibfnamefont {P.}~\bibnamefont {Hopf}}, \bibinfo {author} {\bibfnamefont {G.}~\bibnamefont {Su{\'{a}}rez}}, \bibinfo {author} {\bibfnamefont {M.}~\bibnamefont {Gali}}, \bibinfo {author} {\bibfnamefont {J.}~\bibnamefont {Lishman}}, \bibinfo {author} {\bibfnamefont {R.}~\bibnamefont {Gadhvi}}, \bibinfo {author} {\bibfnamefont {R.}~\bibnamefont {Agarwal}}, \bibinfo {author} {\bibfnamefont {A.}~\bibnamefont {Galicia}}, \bibinfo {author} {\bibfnamefont {N.}~\bibnamefont {Shammah}}, \bibinfo {author} {\bibfnamefont {P.}~\bibnamefont {Nation}}, \bibinfo {author} {\bibfnamefont {J.~R.}\ \bibnamefont {Johansson}}, \bibinfo {author} {\bibfnamefont {S.}~\bibnamefont {Ahmed}}, \bibinfo {author} {\bibfnamefont {S.}~\bibnamefont
  {Cross}}, \bibinfo {author} {\bibfnamefont {A.}~\bibnamefont {Pitchford}},\ and\ \bibinfo {author} {\bibfnamefont {F.}~\bibnamefont {Nori}},\ }\bibfield  {title} {\bibinfo {title} {{QuTiP 5: The Quantum Toolbox in Python}},\ }\href {http://arxiv.org/abs/2412.04705} {\bibfield  {journal} {\bibinfo  {journal} {arXiv}\ ,\ \bibinfo {pages} {2412.04705}} (\bibinfo {year} {2024})}\BibitemShut {NoStop}%
\bibitem [{\citenamefont {Hastrup}\ and\ \citenamefont {Andersen}(2021)}]{Hastrup2021}%
  \BibitemOpen
  \bibfield  {author} {\bibinfo {author} {\bibfnamefont {J.}~\bibnamefont {Hastrup}}\ and\ \bibinfo {author} {\bibfnamefont {U.~L.}\ \bibnamefont {Andersen}},\ }\bibfield  {title} {\bibinfo {title} {{Improved readout of qubit-coupled Gottesman-Kitaev-Preskill states}},\ }\href@noop {} {\bibfield  {journal} {\bibinfo  {journal} {Quantum Science and Technology}\ }\textbf {\bibinfo {volume} {6}},\ \bibinfo {pages} {035016} (\bibinfo {year} {2021})}\BibitemShut {NoStop}%
\bibitem [{\citenamefont {Zak}(1966)}]{Zak1966}%
  \BibitemOpen
  \bibfield  {author} {\bibinfo {author} {\bibfnamefont {J.}~\bibnamefont {Zak}},\ }\bibfield  {title} {\bibinfo {title} {{Dynamics of Electrons in Solids in External Fields}},\ }\href@noop {} {\bibfield  {journal} {\bibinfo  {journal} {Physical Review}\ }\textbf {\bibinfo {volume} {168}},\ \bibinfo {pages} {686} (\bibinfo {year} {1966})}\BibitemShut {NoStop}%
\bibitem [{Note3()}]{Note3}%
  \BibitemOpen
  \bibinfo {note} {Reference~\cite {Royer2020} adopted a different convention $\protect \hat {b}_t \to \protect \frac {\protect \hat {\sigma }_x +i\protect \hat {\sigma }_y}{\protect \sqrt {2\delta t}}$}\BibitemShut {NoStop}%
\bibitem [{Note4()}]{Note4}%
  \BibitemOpen
  \bibinfo {note} {We assign $\mathinner {|{{\protect \rm sq}^\pm _{\alpha , r}}\rangle }$ to the logical $\mathinner {|{\pm }\rangle }$ state rather than $\mathinner {|{0}\rangle }_L, \mathinner {|{1}\rangle }_L$ states, following the notation in Ref.~\cite {Xu2023}.}\BibitemShut {Stop}%
\bibitem [{\citenamefont {Takesaki}(1979)}]{Takesaki1979}%
  \BibitemOpen
  \bibfield  {author} {\bibinfo {author} {\bibfnamefont {M.}~\bibnamefont {Takesaki}},\ }\href {https://doi.org/10.1007/978-1-4612-6188-9} {\emph {\bibinfo {title} {Theory of Operator Algebras I}}}\ (\bibinfo  {publisher} {Springer New York},\ \bibinfo {year} {1979})\BibitemShut {NoStop}%
\end{thebibliography}%

\appendix
\newpage
\onecolumngrid
\begin{center}
    {\bf\large End Matter}
\end{center}
\vspace{1cm}
\twocolumngrid
In this End Matter, we describe two operations necessary for universal computation.

\begin{figure}[htbp]
\centering
\scalebox{1.0}{%
\Qcircuit @C=0.8em @R=.7em {
  \lstick{\hspace{-0.5em}\ket 0} 
    & \qw
    & \gate{\h D(2\sqrt 2\alpha)}
    & \qw
    & \gate{\h D(\frac{i\pi}{4\alpha})} 
    & \qw
    & \rstick{\ket{{\rm cat}^+_{\alpha}}}\\
  \lstick{\hspace{-0.5em}\ket{0}} 
    & \gate{H}  
    & \ctrl{-1}
    & \meterB{X}
    & \control \cw \cwx
}}
\caption{A circuit for generating a cat state.}
\label{figEM:cat circuit}
\end{figure}
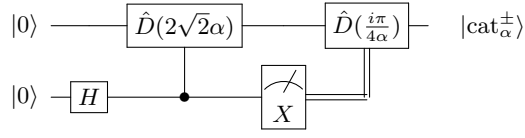

\begin{figure}[htbp]
    \centering
    \includegraphics[width=\linewidth]{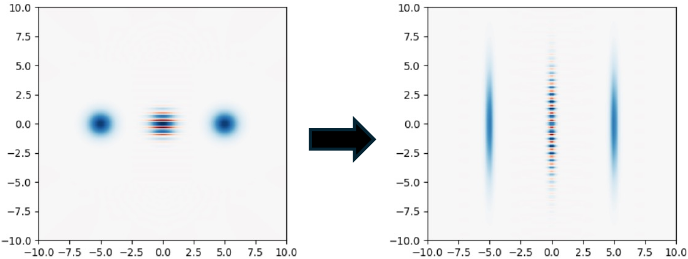}
    \caption{By applying the QEC protocol repeatedly, the cat state is driven to the squeezed cat state.}
    \label{figEM:StatePreparation}
\end{figure}

\textit{\textbf{State preparation}}: $\mathcal P_{\ket +}$---
To generate a $\ket{+}$ state in the SC code, we first generate a cat state by a circuit shown in Fig.~\ref{figEM:cat circuit}.
Then, we obtain the SC state by applying our QEC protocol multiple times, as demonstrated in Fig.~\ref{figEM:StatePreparation}.

\textit{\textbf{Logical $Z$ readout}}: $\calM_Z$---
In this section, we propose a protocol for measuring the logical $Z$ operator.
We first heuristically derive a circuit to measure the logical $Z$ operator, and then we numerically verify that the error scaling is indeed improved compared with that obtained using a na{\"i}ve Hadamard test of logical $Z$ operator.

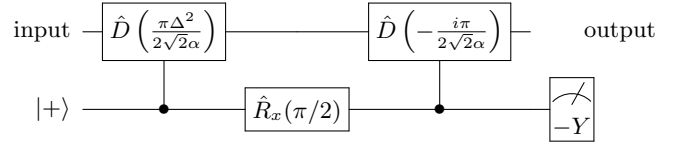
\begin{figure}[htbp]
\centering
\scalebox{1.0}{%
\Qcircuit @C=0.8em @R=.7em {
  \lstick{\hspace{-0.5em}\rm input} 
    & \gate{\h D\kakko{\frac{\pi \Delta^2}{2\sqrt{2} \alpha}}} 
    & \qw 
    & \gate{\h D\kakko{-\frac{i\pi}{2\sqrt{2} \alpha}}}
    & \qw 
    & \rstick{\rm output}\\
  \lstick{\hspace{-0.5em}\ket{+}} 
    & \ctrl{-1}   
    & \gate{\h{R}_x(\pi / 2)} 
    & \ctrl{-1}
    & \qw 
    & \meterB{-Y}
}}
\caption{An improved measurement circuit of $\h Z_L$ for the squeezed cat code, corresponding to the trim-like Trotterization~\eqref{eq:MeasTrim}.}
\label{fig:LogZ circuit}
\end{figure}

For the infinitely-squeezed cat state, the logical $Z$ operator can be chosen to be $\h Z_0 = -i\h D(i\frac{\pi}{4\alpha})$.
Therefore, a na{\"i}ve way to measure $\h Z_0$ is to perform the Hadamard test utilizing the conditional displacement operator $\expo{i\frac{\pi}{4\sqrt 2\alpha}\h x\otimes \h\sigma_x}$.
In the case of the finitely-squeezed cat state, the logical $Z$ operator is modified as
\e{
\h Z_\Delta = \h E_\Delta \h Z_0 \h E_\Delta^{-1} = \expo{i\frac{\pi\Delta}{2\alpha}d'_\Delta},
}
where
$\h d'_\Delta = \frac{1}{\sqrt{2}} \kakko{\frac{\h{x}_{[4\sqrt{2}\alpha]}}{\Delta} + i \Delta \h{p}}$ is a non-Hermitian operator. 
The interaction term $\h d'_\Delta\otimes \h\sigma_x$ is also non-Hermitian, so we slightly modify it as
\e{
\h d'_\Delta\otimes \h\sigma_x &= \h d'_\Delta\otimes (\h\sigma_+ + \h\sigma_-) \nonumber \\
&\simeq \h d'_\Delta\otimes \h\sigma_- +  \h d'^\dagger_\Delta\otimes \h\sigma_+,
}
which is Hermitian. The unitary operator realizing the interaction is then given by
\e{
\h U = \expo{-i\frac{\pi}{4\sqrt 2\alpha}\kakko{\h x_{[4\sqrt 2\alpha]}\otimes\h\sigma_x + \Delta^2\h p\otimes\h\sigma_y}}. \label{eq:MeasUnit}
}
We Trotterize this interaction unitary operator, imposing the constraint that the replacement of modular operator $\h x_{[4\sqrt 2\alpha]}$ with $\h x$ results in only a trivial operation on the qubit, i.e., a global phase factor.
Noting that the $\h p\otimes \sigma_y$ term commutes with $\h\sigma_y$ to be measured, we see that it does not affect the measurement result if placed last. Therefore, we place this term first, and obtain the trim-type decomposition as
\e{
\h U^{(T)} = \expo{-i\frac{\pi}{4\sqrt 2\alpha} \h x\otimes \sigma_x} \expo{-i\frac{\pi\Delta^2}{4\sqrt 2\alpha} \h p\otimes \sigma_y}.  \label{eq:MeasTrim}
}
An equivalent circuit using the conditional displacement operator and a qubit rotation is given in Fig.~\ref{fig:LogZ circuit}.
We note that a similar circuit has been proposed for measuring logical Pauli operators for the GKP code~\cite{Hastrup2021,Royer2020}.

Finally, we numerically confirm the effectiveness of our improved circuit for measuring $\h Z_L$.
As possible realizations of the measurement of $Z_L$, we consider the na{\"i}ve Hadamard test for $Z_0$, measurement circuits corresponding to several types of Trotterization of Eq.~\eqref{eq:MeasUnit} (sharpen, trim, BsB, and sBs), and the Homodyne measurement.
We define the error probability $p_{\rm err}:=(p(1|0)+p(0|1))/2$, where $p(1|0) (p(0|1))$ is the probability of obtaining the measurement outcome 1(0) where the true state is $\ket 0_L (\ket 1_L)$.

\begin{figure}[htbp]
    \centering
    \includegraphics[width=\linewidth]{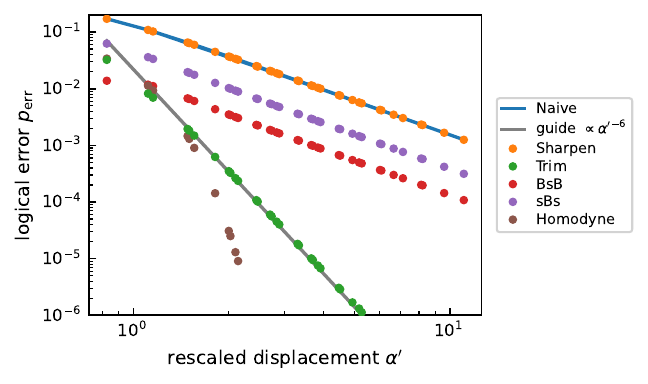}
    \caption{Measurement error $p_{\rm err}=(p(1|0)+p(0|1))/2$ in the logical measurement of $\h{Z}_L$ with different protocols. For the na{\"i}ve protocol, the logical error scales as $p_{\rm err}\propto \alpha'^{-2}$, while $p_{\rm err}\propto \alpha'^{-6}$ for the trim-like circuit in Fig.~\ref{fig:LogZ circuit}.}
    \label{fig:LogicalMeasurement}
\end{figure}

In Fig.~\ref{fig:LogicalMeasurement}, we plot the error probability for different measurement protocols against the rescaled displacement $\alpha' = \alpha  e^r$.
We confirm that the error probability in the trim-like circuit measurement scale as $\alpha'^{-6}$, while the error probability obtained using other circuit-based protocols scales as $\alpha'^{-2}$, thereby showing the cubic improvement.

While the error of the proposed measurement scheme does not reach the fundamental limit set by the Helstrom bound, it achieves a significant improvement in error scaling—from $\alpha'^{-2}$ to $\alpha'^{-6}$—compared with a na\"{i}ve measurement based solely on translational symmetry. Moreover, the proposed scheme can be particularly advantageous in systems such as circuit QED architectures, where homodyne detection is challenging.

\clearpage

\onecolumngrid
\renewcommand{\thetheorem}{S\arabic{theorem}}
\renewcommand{\thelemma}{S\arabic{lemma}}
\renewcommand{\thecorollary}{S\arabic{corollary}}
\renewcommand{\thesection}{S\arabic{section}}
\setcounter{equation}{0}
\setcounter{figure}{0}
\setcounter{theorem}{0}
\setcounter{lemma}{0}
\setcounter{page}{1}
\setcounter{section}{0}
\counterwithout{equation}{section}
\renewcommand{\theequation}{S.\arabic{equation}}
\renewcommand{\thefigure}{S.\arabic{figure}}

\begin{center}
{\large \bf Supplemental Material:  
Exploiting Translational Symmetry for Quantum Computing 
with Squeezed Cat Qubits}\\
\vspace*{0.3cm}
Tomohiro Shitara$^{1}$, Gabriel Mintzer$^{2}$, Yuuki Tokunaga$^{1}$, and Suguru Endo$^{1}$\\
\vspace*{0.1cm}

$^{1}$NTT Computer and Data Science Laboratories, NTT Corporation, 3-9-11 Midori-cho, Musashino-shi, Tokyo 180-8585, Japan

$^{2}$MIT Department of Electrical Engineering and Computer Science, 50 Vassar St, Cambridge, MA 02139, USA
\end{center}

\tableofcontents

\bigskip

\section{Squeezed cat state}\label{secSM:squeezed cat}
In this section, we discuss how we can obtain the finitely-squeezed squeezed cat (SC) state $\ket{{\rm sq}^\pm_{\alpha, r}}$ by applying the envelope operator $ e^{-\Delta^2 \h{p}^2/2}$ to the ideal infinitely-squeezed SC state $\ket{{\rm sq}^\pm_{\alpha,    r = \infty}}$.
We carry out our analysis in the wavefunction representation, either in position or momentum space, for convenience.

\subsection{Wavefunction representations of operators}
For a quantum state $\ket{\psi}$, the wavefunction in the position basis and that in the momentum basis are defined by
\e{
\psi(x) &\coloneq \braket{x | \psi},\\
\tilde{\psi}(p) &\coloneq \braket{p | \psi},
}
respectively. The position and momentum eigenstates $\ket{x}, \ket{p}$ are related as
\e{
\braket{x | p} = \frac{1}{\sqrt{2 \pi}}  e^{i p x}.
}
As such, $\psi(x)$ and $\tilde{\psi}(p)$ are related via the Fourier transformation as
\begin{align}
\psi(x) = \braket{x | \psi}= \int \intd p \  \braket{x | p}\braket{p | \psi} 
= \frac{1}{\sqrt{2 \pi}} \int \intd p \  e^{i p x} \tilde{\psi}(p),
\end{align}
and in a similar manner, we obtain
\begin{align}
\tilde{\psi}(p) = \frac{1}{\sqrt{2 \pi}} \int \intd x \  e^{-i p x} \psi(x).
\end{align}

The displacement operator $\h{D}(\alpha) \coloneq \expo{\alpha \h{a}^\dagger - \alpha^* \h{a}}$ acts on the quadrature operators as
\e{
\h{D}^\dagger(\alpha) \h{a} \h{D}(\alpha) &= \h{a} + \alpha,\\
\h{D}^\dagger(\alpha) \h{a}^\dagger \h{D}(\alpha) &= \h{a}^\dagger + \alpha^*,\\
\h{D}^\dagger(\alpha) \h{x} \h{D}(\alpha) &= \h{x} + \sqrt{2}\re[\alpha],\\
\h{D}^\dagger(\alpha) \h{p} \h{D}(\alpha) &= \h{p} + \sqrt{2}\im[\alpha].
}
Similarly, the squeezing operator $\h{S}(z) \coloneq \expo{\frac{1}{2} (z^* \h{a}^2 - z {\left( \h{a}^\dagger \right)}^2)}$ acts as
\e{
\h{S}^\dagger(r) \h{a} \h{S}(r) &= \h{a} \cosh{r} - \h{a}^\dagger \sinh{r},\\
\h{S}^\dagger(r) \h{a}^\dagger \h{S}(r) &= \h{a}^\dagger \cosh{r} - \h{a} \sinh{r},\\
\h{S}^\dagger(r) \h{x} \h{S}(r) &=  e^{-r} \h{x},\\
\h{S}^\dagger(r) \h{p} \h{S}(r) &=  e^{r} \h{p}.
}
for a real squeezing parameter $r \in \mathbb{R}$.
Next, we find the $x$-representation of $\h{D}(\alpha)$ and $\h{S}(r)$ for $\alpha \in \mathbb{R}$.
We see that $\h{D}(\alpha) \ket{x}$ is also an eigenstate of $\h{x}$, as
\begin{align}
\h{x} \h{D}(\alpha)\ket x 
= \h{D}(\alpha)\h{D}^\dagger(\alpha) \h{x} \h{D}(\alpha)\ket{x} 
= \h{D}(\alpha)(\h{x} + \sqrt{2} \alpha) \ket{x} 
= (x + \sqrt{2} \alpha) \h{D}(\alpha) \ket{x}.
\end{align}
This implies that $\h{D}(\alpha)\ket{x} = \ket{x + \sqrt{2} \alpha}$, and therefore,
\e{
\bra{x} \h{D}(\alpha) 
= \kakko{\h D(\alpha)^\dagger  \ket{x}}^\dagger
=\kakko{\h D(-\alpha)  \ket{x}}^\dagger
= \kakko{\ket{x - \sqrt{2}\alpha}}^\dagger
= \bra{x - \sqrt{2} \alpha}.
}
For $\h{S}(r)$, we see that
\begin{align}
\h{x} \h{S}(r)\ket{x} 
= \h{S}(r) \h{S}^\dagger(r) \h{x} \h{S}(r) \ket{x} 
= \h{S}(r)  e^{-r} \h{x} \ket{x} 
=  e^{-r} x \h{S}(r) \ket{x},
\end{align}
implying that $\h{S}(r) \ket{x} = C \ket{ e^{-r} x}$, where $C$ is a normalization constant. To identify the value of $C$, we calculate the inner product as
\begin{align}
\delta(x - x') 
= \braket{x | x'}
= \braket{x | \h{S}^\dagger(r) \h{S}(r) | x'} 
= C^2 \braket{ e^{-r} x |  e^{-r} x'} 
= C^2 \delta ( e^{-r} (x - x')) 
= C^2  e^{r} \delta(x - x').
\end{align}
We obtain $C =  e^{-r / 2}$, and hence,
\e{
\h S(r)\ket x =  e^{-r / 2} \ket{ e^{-r} x}.
}
Therefore, we find
\begin{align}
\bra x \h{S}(r) 
= (\h{S}^\dagger(r) \ket{x})^\dagger 
= (\h{S}(-r) \ket{x})^\dagger 
= ( e^{r / 2} \ket{ e^r x})^\dagger 
=  e^{r / 2}\bra{ e^r x}.
\end{align}

\subsection{Envelope operator for displaced squeezed state}
We first derive the wavefunction of the displaced squeezed state $\ket{\alpha,r}\coloneq\h{D}(\alpha) \h{S}(r) \ket{\vac}$, based on the basic relations derived above.
Noting that the wavefunction of the vacuum state is given by
\e{
\braket{x | \vac} = \frac{1}{\pi^{1 / 4}}  e^{-x^2 / 2},
}
we obtain the wavefunction of the displaced squeezed state in the position basis as
\begin{align}
\psi_{\rm ds}(x; \alpha, r)
&=\braket{x | \h{D}(\alpha) \h{S}(r) | \vac} 
= \braket{x - \sqrt{2} \alpha | \h{S}(r)|\vac} \nonumber\\
&= e^{r / 2}\braket{ e^r (x - \sqrt{2} \alpha) | \vac}
=\frac{ e^{r / 2}}{\pi^{1 / 4}} e^{-\frac{ e^{2 r}(x - \sqrt{2} \alpha)^2}{2}}.
\end{align}
The wavefunction in the momentum basis can also be obtained via the Fourier transformation as
\e{
\tilde\psi_{\rm ds}(p; \alpha, r) 
= \frac{1}{\sqrt{2 \pi}} \int \intd x \  e^{-i p x} \psi_{\rm ds}(x; \alpha, r)
=\frac{ e^{-r / 2}}{\pi^{1 / 4}} e^{-\frac{e^{-2r}p^2}{2  } - \sqrt{2} i \alpha p}.
}
Therefore, if we apply an envelope operator $\h E_\Delta =  e^{-\Delta^2\h p^2/2}$ to the displaced squeezed state, it is deformed as
\e{
\h E_\Delta\tilde\psi_{\rm ds}(p; \alpha, r) = \frac{ e^{-r / 2}}{\pi^{1 / 4}}\expo{-\frac{ e^{-2r} + \Delta^2}{2} p^2 - \sqrt{2} i \alpha p} 
=  e^{(r' - r) / 2} \tilde{\psi}_{\rm ds}(p; \alpha, r'),  \label{eq:displaced Fock envelope}
}
where the new squeezing parameter $r'$ satisfies
\e{
 e^{-2 r'} =  e^{-2 r} + \Delta^2 \Leftrightarrow r' = -\frac{1}{2} \log\kakko{ e^{-2r} + \Delta^2}. \label{eq:displaced Fock envelope2}
}
This means that if we apply the envelope operator to the SC state, then the squeezing level decreases as $r \to r' (< r)$.
In particular, when the initial state is infinitely squeezed, i.e., $r=\infty$, the final squeezing level $r'$ and the cutoff parameter $\Delta$ are connected via the simple relation
\e{
r' = -\log\Delta.
}

\subsection{Stabilizer operator for squeezed cat state}
The SC state is defined as
\e{
\ket{{\rm sq}^\pm_{\alpha, r}} \coloneq \frac{1}{\sqrt{\calN_{0}^{\pm}}}\kakko{\ket{\alpha, r} \pm \ket{-\alpha,    r}},
}
where 
\e{
\calN_{0}^{\pm}=\kakko{\bra{\alpha, r} \pm \bra{-\alpha, r}}\kakko{\ket{\alpha, r}\pm \ket{-\alpha, r}}
}
is the normalization factor. 
Since the finitely-squeezed coherent state is obtained by applying the envelope operator to the infinitely-squeezed coherent state, the finitely-squeezed cat state is obtained from the infinitely-squeezed cat state by the same transformation:
\e{
\ket{{\rm sq}^\pm_{\alpha, r}}\propto \h{E}_\Delta \ket{{\rm sq}^\pm_{\alpha, r = \infty}} \label{eqSM:squeezingSC}
}
with $r = -\log\Delta$.
Given that the infinitely-squeezed cat state is stabilized by $\h{T}_0 = \h{D}(i \pi / \alpha) =  e^{\sqrt{2} i \pi \h x / \alpha}$, i.e., $\h{T}_0\ket{{\rm sq}^\pm_{\alpha, r = \infty}} =\ket{{\rm sq}^\pm_{\alpha, r = \infty}}$, the finitely-squeezed cat state is stabilized by 
\e{
\h{T}_\Delta &= \h{E}_\Delta \h{T}_0 \h{E}_\Delta^{-1} 
= \expo{\sqrt{2} i \pi (\h E_\Delta \h x \h E_\Delta^{-1}) / \alpha}   
= \expo{\frac{\sqrt{2} \pi}{\alpha}(i \h x - \Delta^2 \h p)}.
}
In the last equality, we have used $\h E_\Delta \h x \h E_\Delta^{-1} = \h x + i \Delta^2 \h p$, which can be shown by applying the Campbell identity
\e{
 e^{\h{A}} \h{B}  e^{-\h{A}} 
&=\sum_{n = 0}^\infty \frac{1}{n!}{\rm ad}_{\h{A}}^n (\h{B}) \nonumber \\
&= \h{B} + [\h{A},    \h{B}] + \frac{1}{2!}[\h{A},    [\h{A},    \h{B}]] + \frac{1}{3!}[\h{A},    [\h{A},    [\h{A},    \h{B}]]] + \cdots,
}
where ${\rm ad}_{\h{A}} (\h{B}) = [\h{A}, \h{B}] = \h{A} \h{B} - \h{B} \h{A}$ is the adjoint superoperator.


\section{Autonomous error-correction for squeezed cat code}\label{sec:dissipative-error-correction}
As pointed out in Refs.~\cite{Schlegel2022,Endo2024}, the ideal, infinitely-squeezed cat state $\ket{{\rm sq}^\pm_{\alpha,    r = \infty}}$ has a discrete translational symmetry and hence is stabilized by $-\h{D} \left( \frac{i \pi}{2 \alpha} \right)$, that is,
\e{
-\h{D}\kakko{\frac{i \pi}{2 \alpha}} \ket{{\rm sq}^\pm_{\alpha, r = \infty}} = \ket{{\rm sq}^\pm_{\alpha, r = \infty}}.
}
The infinitely-squeezed cat state is also stabilized by $\h{T}_0 = \h{D}\left( \frac{i \pi}{\alpha} \right) = \kakko{-\h{D} \left( \frac{i \pi}{2 \alpha} \right)}^2$.
The choice of $\h{T}_0$ rather than $-\h{D} \left( \frac{i \pi}{2 \alpha} \right)$ as the stabilizer is made in order to avoid the minus sign.
As shown in Eq.~\eqref{eqSM:squeezingSC}, the finitely-squeezed cat state is obtained by applying the envelope operator $\h{E}_\Delta =  e^{-\Delta^2 \h{p}^2 / 2}$ to the infinitely-squeezed cat state.
Therefore, the finitely-squeezed cat state is stabilized by
\e{
\h{T}_\Delta = \h{E}_\Delta \h{T}_0 \h{E}_\Delta^{-1} =  e^{\frac{\sqrt{2} \pi}{\alpha}(i \h{x} - \Delta^2 \h{p})}.
}
with $\Delta=\ex^{-r}$.

The fact that the finitely-squeezed cat state is stabilized by $\h{T}_\Delta =  e^{\frac{\sqrt{2} \pi}{\alpha}(i \h{x} - \Delta^2 \h{p})}$ gives us an insight regarding how to stabilize the SC state by dissipation.
We define the dissipator $\h{d}_\Delta$ by
\e{
\h{d}_\Delta = \frac{1}{\sqrt{2}} \kakko{\frac{\h{x}_{[\sqrt{2} \alpha]}}{\Delta} + i \Delta \h{p}}.
}
Here, $\h{x}_{[\sqrt{2} \alpha]}$ is the modular position operator, or the position operator in the Zak basis~\cite{Zak1966,Zak1967}, satisfying $\h{x}_{[\sqrt{2} \alpha]} = \h{x} \ {\rm mod}\ \sqrt{2} \alpha$ and $-\alpha / \sqrt{2} < \h{x}_{[\sqrt{2} \alpha]} \le \alpha / \sqrt{2}$. 
We note that the dissipator is chosen so that $\h{d}_\Delta \propto \log \h{T}_\Delta$ and is normalized in the sense that it behaves like an annihilation operator if we neglect the modularity, that is, $\lim_{\alpha \to \infty}[\h{d}_\Delta, \h{d}_\Delta^\dagger] = \h{I}$.
For this choice of the dissipator, the stability condition is equivalent to annihilating the state by the dissipator, i.e., 
\e{
\h{T}_\Delta \ket{\psi} = \ket{\psi} \Leftrightarrow \h{d}_\Delta \ket{\psi} = 0.
}

Once the dissipator necessary for stabilizing the SC state is identified, we can construct a circuit that realizes the dissipation with an ancillary qubit  by following a similar procedure to that utilized in Ref.~\cite{Royer2020}.

To induce the dissipation $\calD[\h d_\Delta]$ on the bosonic system, let the system interact with the environmental bosonic modes with interaction Hamiltonian described by
\e{
\h H_{\rm int}(t)=\sqrt\Gamma (\h d_\Delta \h b_t^\dagger + \h d_\Delta^\dagger \h b_t).
}
Here, $\h b_t$ is the bosonic annihilation operator satisfying $[\h b_t, \h b_{t'}^\dagger]=\delta(t-t')$.
The state of the environment is set to be the vacuum, satisfying $\braket{\h b_t^\dagger \h b_t}=0$.

To mimic these dynamics with a quantum circuit, we first discretize the dynamics and the bosonic modes with $\delta t$ being the discretized time step, and then we replace the bosonic annihilation operator with the qubit lowering operator ~\footnote{Reference~\cite{Royer2020} adopted a different convention $\h b_t \to \frac{\h\sigma_x +i\h\sigma_y}{\sqrt{2\delta t}}$} as
\e{
\h b_t \to \frac{\h\sigma_x +i\h\sigma_y}{2\sqrt{\delta t}}.
}
The state of the qubit is reset to its ground state $\ket 0$ after each time evolution of duration $\delta t$ to ensure the independence of each time step.
The replacement of the bosonic environment with a qubit is justified if the number of excitations generated during each time step is much smaller than one, which is expected to hold for sufficiently small $\Gamma \delta t$.
The unitary operator applied during the time interval $[t,t+\delta t]$ is then given by
\e{
\h U&=\expo{-i\h H_{\rm int}\delta t} \nonumber\\
&=\expo{-i\sqrt{\frac{\Gamma\delta t}{2}}(\h x_{[\sqrt 2\alpha]}\h\sigma_x /\Delta + \Delta \h p\h\sigma_y)}.
}
Using the Trotterization of the lowest order, we obtain the sharpen and trim processes $\h U'_{\rm S}$ and $\h U'_{\rm T}$ as
\e{
\h U'_{\rm S}&=\expo{-i\sqrt{\frac{\Gamma\delta t}{2}}\Delta \h p\h\sigma_y}\expo{-i\sqrt{\frac{\Gamma\delta t}{2}}\h x_{[\sqrt 2\alpha]}\h\sigma_x /\Delta },\\
\h U'_{\rm T}&=\expo{-i\sqrt{\frac{\Gamma\delta t}{2}}\h x_{[\sqrt 2\alpha]}\h\sigma_x /\Delta }
\expo{-i\sqrt{\frac{\Gamma\delta t}{2}}\Delta \h p\h\sigma_y}.
}
The unitary operator generated by the modular position operator is equivalent to that generated by the position operator up to a global phase factor if the time step $\delta t$ satisfies the modularity condition
\e{
\sqrt{\frac{\Gamma\delta t}{2}}\cdot\frac{\sqrt2\alpha}{\Delta}=\pi,
}
which is equivalent to $\Gamma\delta t=\pi^2 e^{-2r}/\alpha^2$.
Substituting this into the sharpen and trim unitary operators, we obtain
\e{
U^{(\rm ST)} 
=
\begin{cases}
    \expo{-i\frac{\pi\Delta^2}{\sqrt{2} \alpha} \h{p} \otimes \h{\sigma}_y}\expo{-i \frac{\pi}{\sqrt{2} \alpha} \h{x} \otimes \h{\sigma}_x}& \text{(sharpen)} , \\
    \expo{-i\frac{\pi}{\sqrt{2} \alpha} \h{x} \otimes\h{\sigma}_x }\expo{-i \frac{\pi \Delta^2}{\sqrt{2} \alpha} \h{p} \otimes \h{\sigma}_y}& \text{(trim)},
\end{cases}
}

with the ancillary qubit being initialized to $\ket 0$.
Equivalently, the dissipation process can also be realized by the circuits in Fig.~\ref{figSM:ST circuit}.
The conditional displacement operation is defined as $C \h{D}(\beta) = \expo{(\beta a^\dagger - \beta^* a) \h{\sigma}_z / 2 \sqrt{2}}$, and the ancilla rotation is given by $\h{R}_x(\theta) = \expo{-i \theta \h{\sigma}_x / 2}$.
The comparison between our method and that proposed in Ref.~\cite{Royer2020} is summarized in Table~\ref{tabSM:comparison_SM}.

Analogously, the sBs and BsB unitary operators, which correspond to the second-order Trotter decompositions, can be shown to be 
\e{
U^{(\rm sBs)} 
=
    \expo{-i\frac{\pi\Delta^2}{2\sqrt{2} \alpha} \h{p} \otimes \h{\sigma}_y}\expo{-i \frac{\pi}{\sqrt{2} \alpha} \h{x} \otimes \h{\sigma}_x}\expo{-i\frac{\pi\Delta^2}{2\sqrt{2} \alpha} \h{p} \otimes \h{\sigma}_y},\\
    U^{(\rm BsB)} 
=
    \expo{-i \frac{\pi}{\sqrt{2} \alpha} \h{x} \otimes \h{\sigma}_x}\expo{-i\frac{\sqrt{2}\pi\Delta^2}{ \alpha} \h{p} \otimes \h{\sigma}_y}\expo{-i \frac{\pi}{\sqrt{2} \alpha} \h{x} \otimes \h{\sigma}_x}.
}

\begin{figure}[htbp]
    \centering
    
    {\textbf{Sharpen circuit}} \par\vspace{0.5em}
    \begin{minipage}{0.9\linewidth}
        \centering
        \Qcircuit @C=1em @R=.7em {
            \lstick{\rm input}
                & \gate{\h D\kakko{-\frac{i\sqrt 2\pi}{\alpha}}}
                & \qw
                & \gate{\h D\kakko{\frac{\sqrt 2\pi \Delta^2}{\alpha}}}
                & \qw
                & \rstick{\rm output}
            \\
            \lstick{\ket{+}}
                & \ctrl{-1}
                & \gate{\h{R}_x^\dagger(\pi/2)}
                & \ctrl{-1}
                & \qw
                & \rstick{\rm reset}
        }
    \end{minipage}

    \vspace{7mm}

    {\textbf{Trim circuit}} \par\vspace{0.5em}
    \begin{minipage}{0.9\linewidth}
        \centering
        \Qcircuit @C=1em @R=.7em {
            \lstick{\rm input}
                & \gate{\h D\kakko{\frac{\sqrt 2\pi \Delta^2}{\alpha}}}
                & \qw
                & \gate{\h D\kakko{-\frac{i\sqrt 2\pi}{\alpha}}}
                & \qw
                & \rstick{\rm output}
            \\
            \lstick{\ket{+}}
                & \ctrl{-1}
                & \gate{\h{R}_x(\pi/2)}
                & \ctrl{-1}
                & \qw
                & \rstick{\rm reset}
        }
    \end{minipage}

    \vspace{7mm}

    {\textbf{sBs circuit}} \par\vspace{0.5em}
    \begin{minipage}{0.9\linewidth}
        \centering
        \Qcircuit @C=1em @R=.7em {
            \lstick{\rm input}
                & \gate{\h D\kakko{\frac{\sqrt 2\pi \Delta^2}{\alpha}}}
                & \qw
                & \gate{\h D\kakko{-\frac{i\sqrt 2\pi}{\alpha}}}
                & \qw
                & \gate{\h D\kakko{\frac{\sqrt 2\pi \Delta^2}{\alpha}}}
                & \qw
                & \rstick{\rm output}
            \\
            \lstick{\ket{+}}
                & \ctrl{-1}
                & \gate{\h{R}_x(\pi/2)}
                & \ctrl{-1}
                & \gate{\h{R}_x^\dagger(\pi/2)}
                & \ctrl{-1}
                & \qw
                & \rstick{\rm reset}
        }
    \end{minipage}

    \vspace{7mm}

    {\textbf{BsB circuit}} \par\vspace{0.5em}
    \begin{minipage}{0.9\linewidth}
        \centering
        \Qcircuit @C=1em @R=.7em {
            \lstick{\rm input}
                & \gate{\h D\kakko{-\frac{i\sqrt 2\pi}{\alpha}}}
                & \qw
                & \gate{\h D\kakko{\frac{\sqrt 2\pi \Delta^2}{\alpha}}}
                & \qw
                & \gate{\h D\kakko{-\frac{i\sqrt 2\pi}{\alpha}}}
                & \qw
                & \rstick{\rm output}
            \\
            \lstick{\ket{+}}
                & \ctrl{-1}
                & \gate{\h{R}_x^\dagger(\pi/2)}
                & \ctrl{-1}
                & \gate{\h{R}_x(\pi/2)}
                & \ctrl{-1}
                & \qw
                & \rstick{\rm reset}
        }
    \end{minipage}

    \caption{The circuits for stabilizing the SC code. They consist of conditional displacements on the composite system and $X$-rotations on the ancillary qubit.}
    \label{figSM:ST circuit}
\end{figure}

\begin{table*}[ht]
    \centering
        \caption{Comparison between our method of stabilizing the SC states and the standard sharpen-trim protocol for stabilizing the GKP states in Ref.~\cite{Royer2020}}
    \begin{tabular}{|c|c|c|}
    \hline
       & Our method & GKP stabilization~\cite{Royer2020} \\ \hline
      ideal stabilizer $\h{T}_0$ & $e^{\sqrt{2} i \pi \h{x} / \alpha}$ & $ e^{i l \h{x}}$ (and $ e^{-i l \h{p}}$) \\
      target state & $\ket{{\rm sq}^\pm_{\alpha,    r}}$ & $E_\Delta\ket{{\rm GKP}}$ \\
      envelope $\h{E}_\Delta$ & $ e^{-\Delta^2 \h{p}^2 / 2}$ & $ e^{-\Delta^2 \h{a}^\dagger \h{a}}$\\
      width $\Delta$ & $\Delta =  e^{-r}$ & modular squeezing parameter $\Delta = \frac{1}{l} \sqrt{-\log\left\lvert\tr{\h{T}_0 \rho}\right\rvert^2}$ \\
      stabilizer $\h{T}_\Delta$ & $\expo{\sqrt{2} i \pi (\h{x} + i \Delta^2 \h{p})/\alpha}$ & $\expo{i l (c_\Delta \h{x} + is_\Delta \h{p})}$ \\
      dissipator $\h{d}_\Delta$ & $\frac{1}{\sqrt{2}}\kakko{\frac{\h{x}_{[\sqrt{2} \alpha]}}{\Delta} + i\Delta \h{p}}$ & $\frac{1}{\sqrt{2}}\kakko{\frac{\h{x}_{[l / (2 c_\Delta)]}}{\sqrt{t_\Delta}} + i\sqrt{t_\Delta} \h{p}} $ \\\hline
    \end{tabular}
    \label{tabSM:comparison_SM}
\end{table*}

\section{Analysis of autonomous error correction based on subsystem decomposition}\label{sec:subsystem-analysis}
In this section, we analytically show that the dissipator $\h{d}_{\Delta}$ leads to autonomous QEC in the logical manifold of SC codes. To analyze the action of the dissipator $\h{d}_\Delta$, we introduce the subsystem decomposition~\cite{Xu2023} of the bosonic Hilbert space tailored for the SC states.

\subsection{Subsystem decomposition}
For $\alpha, r > 0$, the displaced squeezed state/squeezed coherent state can be expressed in two ways, reflecting these two interpretations, as
\e{
\ket{\alpha,    r} = \h{D}(\alpha) \h{S}(r) \ket\vac = \h{S}(r) \h{D}(\alpha') \ket\vac,
}
where $\alpha' = \alpha  e^r$.
We can then consider superpositions of the displaced Fock states as
\e{
\ket{\Phi_{\pm, n}} = \frac{1}{\sqrt{\calN_n^\pm}}\kakko{\h{D}(\alpha') \pm (-1)^n \h{D}(-\alpha')} \ket n,
}
where $\calN_n^\pm$ is the normalization factor.
The lowest-$n$ states $\ket{\Phi_{\pm, 0}}$ are the logical states for the cat code, and these basis states are useful for describing the effective low-energy dynamics for the cat code~\cite{Putterman2022,Chamberland2022}.
To deal with the SC code, we work in the squeezed frame, following Ref.~\cite{Xu2023}---i.e., 
\e{
\ket{\Psi_{\pm, n}} = \h{S}(r) \ket{\Phi_{\pm, n}}.
}
We note that the sign $\pm$ corresponds to the parity---that is, $ e^{i\pi \h{a}^\dagger \h{a}} \ket{\Psi_{\pm, n}} = \pm \ket{\Psi_{\pm, n}}$.
This means that two states with a different parity are orthogonal:
\e{
\braket{\Psi_{\pm, n}|\Psi_{\mp, m}} = 0.
}
However, states with the same parity are generally non-orthogonal, with their overlap typically being $O( e^{-2 \alpha'^2})$.
To construct an orthonormal basis, we perform the Gram-Schmidt orthonormalization procedure from lower $n$ in each parity sector. As such, we obtain the complete orthonormal set on the bosonic Hilbert space denoted by
\e{
\ket{\pm}_L \otimes \ket{\tilde n}_G \simeq \ket{\Psi_{\pm, n}}.
}
The equality is approximate because of the $O( e^{-2 \alpha'^2})$ overlap and the Gram-Schmidt orthonormalization procedure.
Hereafter, we often neglect this approximation and just write $\ket{\pm}_L \otimes \ket{\tilde n}_G = \ket{\Psi_{\pm, n}}$, which is justified in the limit $\alpha' \to \infty$.
Now, the total Hilbert space $\calH$ can be decomposed as $\calH \simeq \calH_L \otimes \calH_{G}$, where $\calH_{L}$ and $\calH_{G}$ are the Hilbert spaces representing the logical and gauge degrees of freedom, respectively.
We note that the logical space of the SC code corresponds to the ``vacuum'' in the gauge mode---i.e., $\ket{{\rm sq}^\pm_{\alpha,    r}} = \ket{\pm}_L \otimes \ket{\tilde 0}_G$.

In the subsystem-decomposition basis, the annihilation and creation  operators can be expressed as~\footnote{We assign $\ket{{\rm sq}^\pm_{\alpha,    r}}$ to the logical $\ket\pm$ state rather than $\ket 0_L, \ket 1_L$ states, following the notation in Ref.~\cite{Xu2023}.}
\e{
\h{a} &\simeq  \h{Z}_L \otimes \kakko{\tilde{a} \cosh{r} - \tilde{a}^\dagger \sinh{r} +\alpha \tilde{I}}, \label{eq:annihilation} \\
\h{a}^\dagger &\simeq \h{Z}_L\otimes \kakko{\tilde a^\dagger \cosh{r} -\tilde a \sinh{r} + \alpha \tilde{I}},
}
respectively.
Here, $\h{Z}_L$ is the logical Pauli $Z$ operator acting on the logical mode, where we retain the hat notation for operators acting on the logical mode, and $\tilde{a}, \tilde{a}^\dagger, \tilde{I}$ are the annihilation, creation, and identity operators acting on the gauge mode, respectively.
This expression provides rich implications on the correctability of the SC code against the photon-loss error. The photon-loss error always induces the logical phase-flip error $\h Z_L$, since the photon-loss process changes the parity of the bosonic mode. At the same time, the gauge mode can also be modified. The first term in Eq.~\eqref{eq:annihilation} has no effects, since it vanishes when applied to the SC state. The second term, which is dominant at sufficiently large squeezing level, adds an excitation to the gauge mode, so its contribution to the error is detectable and correctable. On the other hand, the third term does not modify the gauge mode, so its contribution to the error is undetectable.

Using this correspondence, the quadrature operators can also be expressed as
\e{
\h{x} &= \frac{\h{a} + \h{a}^\dagger}{\sqrt{2}} \simeq \h{Z}_L \otimes \kakko{ e^{-r} \tilde{x} 
 + \sqrt{2} \alpha \tilde{I}},\\
\h{p} &= \frac{\h{a} - \h{a}^\dagger}{\sqrt{2} i} \simeq \h{Z}_L \otimes \kakko{ e^{r} \tilde{p}}.
}
Here, $\tilde x = (\tilde{a} + \tilde{a}^\dagger) / \sqrt{2}$ and $\tilde{p} = (\tilde{a} - \tilde{a}^\dagger) / (\sqrt{2} i)$ are the position and momentum operators on the gauge mode.

\subsection{Subsystem decomposition of the dissipation operator of squeezed cat states} \label{subsec:subsys decomp dissipation}

Since the modular position operator can be expanded using trigonometric functions of the usual position operator, we obtain its expression in the subsystem-decomposition basis as
\e{
\h{x}_{[\sqrt{2} \alpha]} =  \h{Z}_L \otimes  e^{-r} \tilde{x}_{[\sqrt{2} \alpha']}. \label{modular position subsystem}
}
See Appendix~\ref{sec:modular} for more details.

Now, we are ready to discuss the properties of the dissipator $\h{d}_\Delta$.
In the subsystem-decomposition basis, we can show
\e{
\h{d}_\Delta &= \frac{1}{\sqrt{2}}\kakko{\frac{\h{x}_{[\sqrt{2} \alpha]}}{\Delta} + i\Delta \h{p}} \nonumber\\
&\simeq \frac{1}{\sqrt{2}} \h{Z}_L\otimes \kakko{\frac{ e^{-r} \tilde{x}_{[\sqrt{2} \alpha']}}{\Delta} + i\Delta  e^{r}\tilde{p}} \nonumber\\
&= \h{Z}_L \otimes \frac{\tilde{x}_{[\sqrt{2} \alpha']} + i \tilde{p}}{\sqrt{2}} ,
}
where we have used the relation $\Delta = e^{-r}$.
In the limit $\alpha' =  e^r \alpha \to \infty$, the period of the modular position operator diverges and the modularity becomes effectively negligible, so we have
\e{
\h{d}_\Delta \simeq \h{Z}_L \otimes \tilde{a} \quad (\alpha' \to \infty).  \label{eqSM:dissipator}
}
This expression is the same as the one obtained in Ref.~\cite{Xu2023}.
This analysis shows a clear advantage of our dissipative error correction over Refs.~\cite{Schlegel2022,Hillmann2023}, since it corrects the logical phase-flip error caused by the photon loss $\h a \simeq  \h{Z}_L \otimes \kakko{\tilde{a} \cosh{r} - \tilde{a}^\dagger \sinh{r} +\alpha \tilde{I}}$.

We note that the limit $\lim_{\alpha'\to\infty}\tilde x_{\sqrt 2\alpha'}=\tilde x$ is a state-dependent notion---i.e., the limit is in the sense of the strong operator topology, not the uniform operator topology~\cite{Takesaki1979}.
We address this point in the next subsection, where we discuss the limitations of the proposed protocol.
It is also worth noting that our dissipator stabilizes in only one direction and that the steady-state space is expected to be strictly larger than the SC logical space, in contrast with the case analyzed in Ref.~\cite{Xu2023}, where the dissipator is designed so that the steady-state space coincides with the logical space.
Surprisingly, in the limit of $\alpha' \to \infty$, which can be realized in the infinite-squeezing limit ($r \to \infty$) or in the large-amplitude limit ($\alpha \to \infty$) for the cat code ($r = 0$), the periodicity becomes effectively negligible and $\h{d}_\Delta$ dissipates the gauge mode to the vacuum state.

\subsection{Limitations}
\label{secSM:limitation}
\begin{figure}[htbp]
    \centering
    \includegraphics[width=0.5\linewidth]{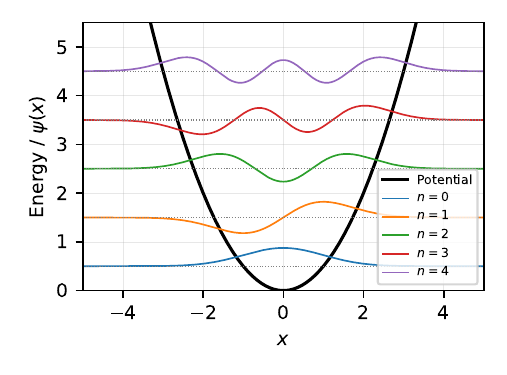}
    \caption{The wavefunction $\psi_n(x)$ of the energy eigenstate of the quantum harmonic oscillator.}
    \label{figSM:Wavefunction}
\end{figure}

Based on the analysis using the subsystem decomposition above, we discuss limitations of the proposed protocol.
First, we note that in deriving the approximate expression in Eq.~\eqref{eq:dissipator} for the dissipator, we have utilized the fact that the action of the modular position operator on the gauge mode can be regarded as equivalent to that of the position operator for sufficiently large $\alpha'$. Since the modular operator and the position operator differ only in their action outside the interval $[-\alpha'/\sqrt 2, \alpha'/\sqrt 2]$, this identification is valid only for states whose wavefunctions vanish outside this region.
The wavefunction of the Fock state $\braket{\tilde x|\tilde n}$ is mainly supported on the classically-allowed region---i.e., $\frac{1}{2}\tilde x^2\le \tilde n + \frac{1}{2}\Leftrightarrow |x|\le \sqrt{2\tilde n + 1}$---as shown in Fig.~\ref{figSM:Wavefunction}. Indeed, the probability of detecting the position outside the classically allowed region vanishes asymptotically as $\sim \tilde n^{-1/3}$~\cite{Jadczyk2015, Paris2015}. Therefore, the modular position in the gauge mode can be approximated to the position operator only for states spanned by the Fock states $\ket{\tilde n}$ with $\sqrt{2\tilde n +1} \ll \alpha'/\sqrt 2$, or simply $\tilde n \ll \alpha'^2/4$.

\begin{figure}[htbp]
    \centering
    \includegraphics[width=0.5\linewidth]{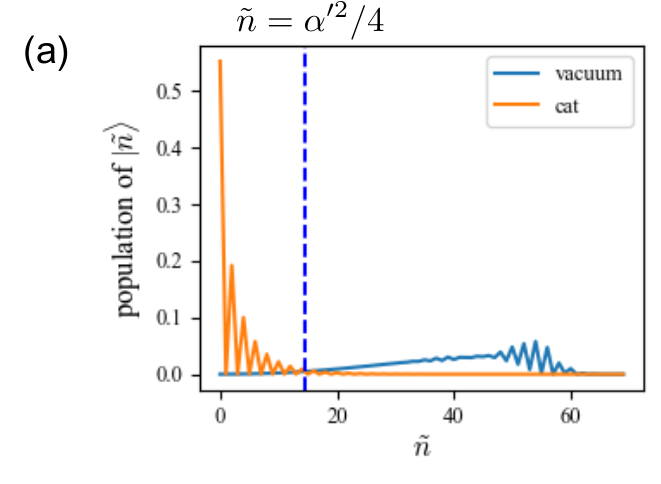}
    \includegraphics[width=0.5\linewidth]{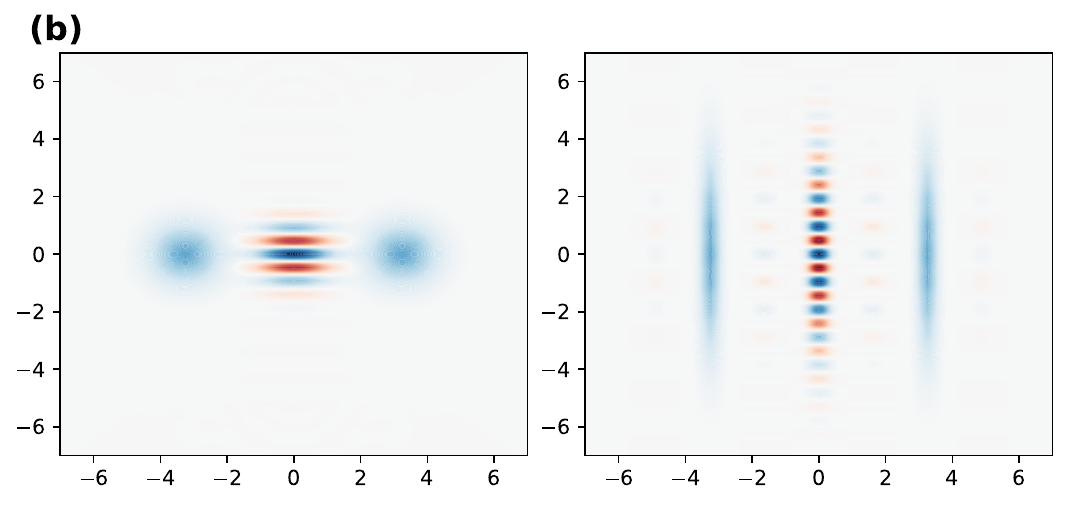}
    \includegraphics[width=0.5\linewidth]{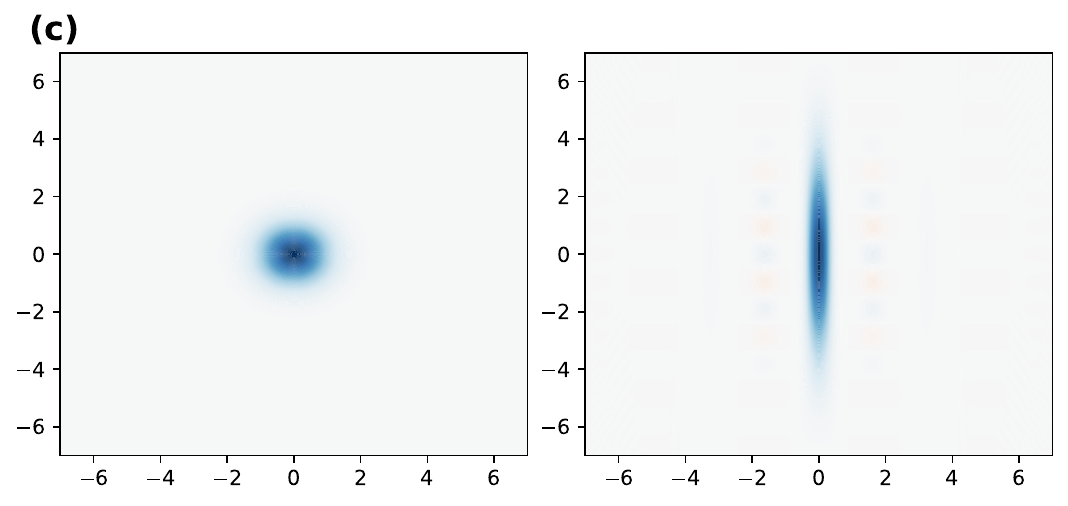}
    \caption{(a)The population of $\ket{\tilde n}_G$ in the gauge mode for the vacuum state $\ket 0$ and the cat state $\ket{{\rm sq}^+_{\alpha,    r = 0}}$. For the cat state, the population decreases in $\tilde n$ and takes small value for $\tilde n\gtrsim \alpha'^2$, while it takes an almost constant value up to $\tilde n\simeq\alpha'^2$ (blue vertical dashed line) for the vacuum state.
    (b) The cat state (left) converges to the SC state (right) after application of 50 cycles of ST.
    (c) The vacuum state (left) converges to a state different from the SC state (right) after application of 50 cycles of ST.
    Parameters are set to be $\alpha=2.3$ and $r=1.2$.}
    \label{figSM:Limitation}
\end{figure}

As a demonstration of this analysis, we consider a preparation of the SC state by applying the sharpen-trim circuits to different types of initial states, namely, the vacuum state and the cat state. Figure~\ref{figSM:Limitation} (a) shows the population of the Fock state $\ket{\tilde n}$ in the gauge mode for these initial states. The population is exponentially decaying in $\tilde n$ for the cat state, while the Fock state population for the vacuum state is distributed almost uniformly on $0 \le \tilde n \le \alpha'^2$. As a result, after applying the sharpen-trim circuit 50 times, the cat state is dissipated to the SC state. However, the vacuum state is not dissipated to the SC state but rather to the squeezed vacuum state, since the sharpen-trim circuit does not drive states $\ket{\tilde n}$ with $\tilde n\gtrsim \alpha'^2$ to the SC state.

It is surprising that in the SDF basis under subsystem decomposition, the cat state $\ket{{\rm sq}^\pm_{\alpha, r = 0}}$ can be written as the product of the logical $+$ state and the squeezed vacuum state in the gauge mode:
\e{
\ket{{\rm cat}^\pm_{\alpha}} \simeq \ket{\pm}_L\otimes \h S(-r)\ket{\tilde 0}_G.
}
More generally, we can show that, in the SDF basis, the SC state with different squeezing parameter can be expressed as the squeezed vacuum in the gauge mode.
To show this, we use Eqs.~\eqref{eq:displaced Fock envelope} and~\eqref{eq:displaced Fock envelope2} to obtain
\e{
\ket{{\rm sq}^\pm_{\alpha, r'}}\propto \h E_\Delta \ket{{\rm sq}^\pm_{\alpha, r}},
}
where $\ex^{-2r'}=\ex^{-2r}+\Delta^2 \Leftrightarrow \Delta^2=\ex^{-2r'}-\ex^{-2r}$ and $\h E_\Delta =\ex^{-\frac{\Delta^2}{2}\h p^2}$.
Therefore, in the SDF basis, we have
\e{
\ket{{\rm sq}^\pm_{\alpha, r'}} &\propto \ex^{-\frac{\Delta^2}{2}\h p^2} \ket{{\rm sq}^\pm_{\alpha, r}} \nonumber\\
& \simeq \ex^{-\frac{\Delta^2}{2}\kakko{\h Z_L\otimes\ex^r\tilde p}^2}\ket{\pm}_L\otimes\ket{\tilde{0}}_G \nonumber\\
&= \ket{\pm}_L\otimes \ex^{-\frac{\Delta^2}{2}\ex^{2r}\tilde{p}^2} \ket{\tilde{0}}_G
}
By using Eqs.~\eqref{eq:displaced Fock envelope} and~\eqref{eq:displaced Fock envelope2} again, we obtain
\e{
\ex^{-\frac{\Delta^2}{2}\ex^{2r}\tilde{p}^2} \ket{\tilde{0}}_G\propto \h S(-r'')\ket{\tilde{0}}_G
}
with $\Delta^2\ex^{2r}+1=\ex^{2r''}$.
Here, $r$, $r'$, and $r''$ are related as $r''=r-r'$, which can be confirmed as
\e{
\ex^{2r''} = \Delta^2 \ex^{2r} +1 = (\ex^{-2r'}-\ex^{-2r})\ex^{2r} +1 = \ex^{2(r-r')}.
}
Therefore, we have $\ket{{\rm sq}^\pm_{\alpha, r'}}=\ket{\pm}_L\otimes \h S_G(r'-r) \ket{\tilde{0}}_G$ and 
\e{
\ket{{\rm cat}^\pm_{\alpha}}=\ket{\pm}_L\otimes S_G(-r) \ket{\tilde{0}}_G.
}
as a special case of $r'=0$.

\subsection{Error analysis}
In this section, we consider two types of autonomous error correction using the subsystem decomposition.
One is given by a jump operator
\e{
\h d_\Delta\simeq \h Z_L\otimes \tilde a,
}
which is discussed in the paper and Ref.~\cite{Xu2023}. Since $\h d_\Delta$ is odd parity, it acts non-trivially on logical space.
The other is given by a jump operator
\e{
\h b^2 -\beta^2\simeq \h I_L\otimes (\tilde a^2 +2\alpha \ex^{-r}\tilde a),
}
which was proposed in Ref.\cite{Hillmann2023}. Since $\h b^2 -\beta^2$ is even parity, it does not change the logical state.

As discussed in the main text, photon-loss error 
\e{
 \h a \simeq \h Z_L\otimes(\tilde a\cosh r -\tilde a^\dagger \sinh r +\alpha\tilde I)
}
induces the logical $Z$ error while partially exciting the gauge mode.
While $\h b^2 -\beta^2$ cannot correct the logical error, $\h d_\Delta$ can partially correct the error induced by the $\h Z_L\otimes \tilde a^\dagger$ term.

On the other hand, the dephasing error is described by the jump operator
\e{
\h a^\dagger \h a \simeq \h I_L\otimes(\tilde a^\dagger\cosh r -\tilde a \sinh r +\alpha\tilde I)(\tilde a\cosh r -\tilde a^\dagger \sinh r +\alpha\tilde I).
}
When the initial state is in the code space as $\ket{\psi}_L\otimes\ket{\tilde 0}_G$, the dephasing error modifies the state as
\e{
\ket{\psi}_L\otimes \kakko{[\alpha^2+\sinh^2 r]\ket{\tilde 0}_G + \alpha\ex^{-r}\ket{\tilde 1}_G - \sqrt{2}\cosh r\sinh r \ket{\tilde 2}_G}.
}
We see that $\h b^2 -\beta^2$ can perfectly correct the dephasing error, while $d_\Delta$ induces an logical error when correcting the $\ket{\psi}_L\otimes \ket{\tilde 1}_G$ term.
However, such an error can be exponentially suppressed by increasing the squeezing parameter $r$.

\section{Modular position operator}\label{sec:modular}
The modular quadrature operator, which is expressed formally written as $\h{x}_{[m]} = \h{x} \mod m$, is defined in the position basis as
\e{
\h{x}_{[m]} = \sum_{k \in \mathbb{Z}}\int_{-m / 2}^{m / 2} \intd x \ x \ket{x + km} \bra{x + km}.
}
It can also be expressed by its Fourier series as
\e{
\h{x}_{[m]}= -\frac{m}{\pi}\sum_{k = 1}^\infty \frac{(-1)^k}{k}\sin\kakko{\frac{2 \pi k \h{x}}{m}}.
}
For a positive constant $c > 0$, one can confirm the formula $(c \h{x})_{[m]} = c \h{x}_{[m / c]}$ as follows:
\e{
(c \h{x})_{[m]} &= -\frac{m}{\pi}\sum_{k = 1}^\infty \frac{(-1)^k}{k} \sin\kakko{\frac{2 \pi k c \h{x}}{m}} \nonumber \\
&= c \cdot \left[-\frac{m / c}{\pi}\sum_{k = 1}^\infty \frac{(-1)^k}{k}\sin\kakko{\frac{2\pi k \h{x}}{m / c}}\right] \nonumber \\
&= c \h{x}_{[m / c]}. \label{dilation}
}
Using this identity, we derive the expression for the modular operator in the subsystem decomposition basis [Eq.~\eqref{modular position subsystem}].
First, we can calculate the trigonometric functions of the position operator and the modular position operator as
\e{
\cos \h{x} &= \sum_{k = 0}^\infty \frac{(-1)^kx^{2k}}{(2k)!} \nonumber \\
&\simeq \sum_{k = 0}^\infty \frac{(-1)^k}{(2k)!}\left[\h{Z}_L \otimes \left( e^{-r} \tilde{x} + \sqrt{2} \alpha \tilde{I}\right)\right]^{2k} \nonumber \\
&= \h{I}_L \otimes \sum_{k = 0}^\infty \frac{(-1)^k\left( e^{-r}\tilde{x} + \sqrt{2} \alpha \tilde{I}\right)^{2k}}{(2k)!} \nonumber \\
&= \h{I}_L \otimes \cos\kakko{ e^{-r} \tilde{x} + \sqrt{2} \alpha \tilde{I}},\\
\sin \h{x} &\simeq \sum_{k = 0}^\infty \frac{(-1)^k}{(2k + 1)!}\left[Z_L \otimes \left( e^{-r} \tilde{x} + \sqrt{2} \alpha \tilde{I}\right)\right]^{2k + 1} \nonumber \\
&= \h{Z}_L \otimes \sin\kakko{ e^{-r} \tilde{x} + \sqrt{2} \alpha \tilde{I}},
}
 and
\e{
\h{x}_{[\sqrt{2} \alpha]}&= -\frac{\sqrt{2} \alpha}{\pi}\sum_{k = 1}^\infty \frac{(-1)^k}{k}\sin\kakko{\frac{2 \pi k \h{x}}{\sqrt{2} \alpha}} \nonumber \\
&\simeq -\frac{\sqrt{2} \alpha}{\pi}\sum_{k = 1}^\infty \frac{(-1)^k}{k} \h{Z}_L \otimes \sin\kakko{\frac{2 \pi k ( e^{-r} \tilde{x} + \sqrt{2} \alpha \tilde{I})}{\sqrt{2} \alpha}} \nonumber \\
&= \h{Z}_L \otimes \left[-\frac{\sqrt{2} \alpha}{\pi}\sum_{k = 1}^\infty \frac{(-1)^k}{k} \sin\kakko{\frac{2 \pi k  e^{-r} \tilde{x}}{\sqrt{2} \alpha}}\right] \nonumber \\
&= \h{Z}_L \otimes \kakko{ e^{-r}\tilde x}_{[\sqrt{2} \alpha]} \nonumber \\
&= \h{Z}_L \otimes  e^{-r} \tilde x_{[\sqrt{2} \alpha  e^{r}]} \nonumber \\
&= \h{Z}_L \otimes  e^{-r} \tilde x_{[\sqrt{2} \alpha']},
}
where we have used Eq.~\eqref{dilation} in the second last line.

\section{Logical operations}
\label{sec:logical-operations}

\subsection{$Z(\theta)$ operation}

In Ref.~\cite{Endo2024}, the logical $Z$ operation is shown to be exactly $-i \h{D}(i \pi / 4 \alpha)$ for the infinite-squeezing-parameter case. For finite squeezing parameter $r$, the gauge photon is excited by the displacement operation $\h{D}(i \theta)~(\theta \in \mathbb{R})$, as shown below. By using the subsystem decomposition, the displacement operator can be written as
\e{
\h{D}(i \theta) =  e^{i 2 \theta \alpha \h{Z}_L} \cdot e^{i \theta e^{-r} \h{Z}_L \otimes (\tilde{a}+ \tilde{a}^\dag)},
\label{eq:logx}
}
which indicates that this operation converges to the $Z$ rotation gate in the logical manifold in the limit of $r \to \infty$ because the second term in Eq.~\eqref{eq:logx} becomes the identity operator. Meanwhile, for finite $r$, the second term induces entanglement between the logical and gauge subsystems for a general state in the code space $\ket{\psi}_C = \ket{\psi}_L \otimes \ket{0}_G $.
For example, taking $\ket{\psi}_L = c_0 \ket{0}_L + c_1 \ket{1}_L $ gives entanglement as follows: 
\e{
\h{D}(i \theta) \ket{\psi}_C =  c_0 e^{i 2 \theta \alpha} \ket{0}_L \otimes \ket{i \theta e^{-r}}_G + c_1 e^{-i 2 \theta \alpha} \ket{1}_L \otimes \ket{-i \theta e^{-r}}_G,
\label{eq:displace}
}
where $\ket{\pm i \theta e^{-r}}$ is the coherent state with amplitude $\pm i \theta e^{-r}$. For sufficiently small $\theta e^{-r}$, we can approximate Eq.~\eqref{eq:displace} as 
\e{
\h{D}(i \theta ) \ket{\psi}_C \sim  (\h{I} + i \theta e^{-r} \h{Z}_L\otimes \tilde{a}^\dag) \ket{\psi (\theta, r)}_C,
}
where $\ket{\psi (\theta, r)}_C = (c_0 e^{i 2 \theta \alpha} \ket{0}_L + c_1 e^{-i 2 \theta \alpha} \ket{1}_L)$. Then, by the application of dissipator $\h{Z}_L \otimes \tilde{a}$, we can correct the errors that occur due to the finite squeezing parameter. We note that this dissipation can be performed by utilizing our dissipative QEC protocol. Thus, we are able to realize the arbitrary rotation $Z$ gate via the repeated application of displacement operators along the momentum axis, followed by our dissipative QEC protocol.

\subsection{$ZZ(\theta)$ operation}
Next, we discuss the two-mode beamsplitter interaction followed by our dissipative QEC protocol. The Hamiltonian for the two-mode beamsplitter interaction reads 
\e{
\h{H}_2 = \frac{\Theta}{2} (\h{a}_1^\dag \h{a}_2 + \h{a}_1 \h{a}_2^\dag ),
}
where $\Theta \in \mathbb{R}$, and $\h{a}_1$ and $\h{a}_2$ are annihilation operators for the first and the second modes. In the subsystem-decomposition basis, we have 
\e{
\h{H}_2 &\sim \Theta \h{Z}_1 \h{Z}_2 \otimes \left[ \alpha^2  +\frac{\alpha e^{-r}}{2} (\tilde{a}_1^\dag + \tilde{a}_2^\dag) -\tilde{a}_1^\dag \tilde{a}_2^\dag \cosh r\sinh r \right],  \label{Eq: multi}
}
where we have extracted the terms that contribute to the first-order dynamics, $(\h{I} - i \h{H}_2 \delta t) \ket{\psi_1}_C \otimes \ket{\psi_2}_C$. Here, $\ket{\psi_k}_C ~ (k = 1, 2)$ denotes the noiseless SC states---i.e., the states with no gauge excitation in the first and the second modes. The first term in Eq.~\eqref{Eq: multi} corresponds to the target logical operation, while the second and third terms lead to the degradation of the logical fidelity. Now, we find 
\e{
e^{- i \h{H}_2 \delta t} \ket{\psi_1}_C \otimes \ket{\psi_2}_C 
\sim \ket{\psi_{12}(\delta t)}_C 
+ \Theta \delta t \h{Z}_1 \h{Z}_2 
 \otimes \bigg\{\frac{\alpha e^{-r}}{2} \left[(\tilde{a}_1^\dag + \tilde{a}_2^\dag) -\mathrm{cosh}(r) \mathrm{sinh}(r) \tilde{a}_1^\dag\tilde{a}_2^\dag \right] \bigg\} 
\ket{\psi_{12}(\delta t)}_C,
}
where $\ket{\psi_{12}(\delta t)}_C = e^{-i \Theta \alpha^2 \h{Z}_1 \h{Z}_2} \ket{\psi_1}_C\otimes \ket{\psi_2}_C$ is the noise-free evolution state. The impact of the second term results in logical error even after applying the dissipative QEC protocol. This is because the errors due to the operators $\h{Z}_1 \h{Z}_2 \tilde{a}_1^\dag$ and $\h{Z}_1 \h{Z}_2 \tilde{a}_2^\dag$ simply become $\h{Z}_2$ and $\h{Z}_1$ after the dissipative QEC protocol. However, the effect of these errors can be suppressed by increasing the squeezing parameter $r$. Meanwhile, the effect of the third term can be canceled with the dissipative QEC protocol for both modes. Therefore, the two-mode operation can be performed by repeated alternating applications of the short-time evolution with the beam-splitter Hamiltonian $\h{H}_2$ and the subsequent dissipative QEC protocol.

\section{Numerical Simulation Results}
\label{sec:code-results}
In this section, we demonstrate the performance of the proposed dissipative QEC protocol and logical operations, by presenting the results from the numerical simulations.
Numerical simulations were performed using the open-source software package QuTiP (Quantum Toolbox in Python)~\cite{Lambert2024}.
We first calculate the matrix element of the annihilation in the subsystem decomposition basis.
Then, all the calculations are done in this basis.

\subsection{Autonomous Quantum Error-Correction}
\label{sec:sharpen-trim}
First, we show how the QEC protocol works against photon-loss errors.
The bosonic system undergoes the photon-loss process $\calE_{\rm loss}$ described by the Gorini–Kossakowski–Sudarshan–Lindblad master equation
\e{
\od{}{\h\rho}{t}=\kappa\h a \h\rho \h a^\dagger -\frac{\kappa}{2}\kakko{\h\rho\h a^\dagger \h a + \h a^\dagger \h a \h\rho} 
}
for the time interval $\kappa t = 0.01$. Then, we apply the QEC circuit $m$ times, written as $\calE_{\rm QEC}^m$.
We evaluate the entanglement fidelity of the total process $\calE_{\rm QEC}^m\circ\calE_{\rm loss}$, defined by
\e{
F_{\rm e}=(\braket{\Phi^{\rm SR} | \calE_{\rm QEC}^m\circ\calE_{\rm loss} \otimes \calI  (\ket{\Phi^{\rm SR}}\bra{\Phi^{\rm SR}})|\Phi^{\rm SR}})^{1/2},
}
where 
\e{
\ket{\Phi^{\rm SR}} \coloneqq \frac{1}{\sqrt{2}}\kakko{\ket{{\rm sq}^+_{\alpha,    r}}_{\rm S}\otimes\ket{{\rm sq}^+_{\alpha,    r}}_{\rm R} + \ket{{\rm sq}^-_{\alpha,    r}}_{\rm S}\otimes\ket{{\rm sq}^-_{\alpha,    r}}_{\rm R}   }
}
is the maximally entangled state on the logical subspaces of the system and a reference system.

\begin{figure}[htbp]
\centering
\begin{subfigure}{0.48\linewidth}
\centering
\includegraphics[width=\linewidth]{figures/Fig_EntanglementFidelityPhotonLossSameCycle.pdf}
\end{subfigure}
\hfill
\begin{subfigure}{0.48\linewidth}
\centering
\includegraphics[width=\linewidth]{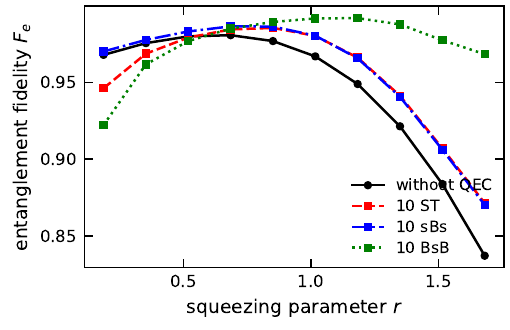}
\end{subfigure}

\caption{Entanglement fidelity against squeezing parameter after photon-loss noise (left) and dephasing noise (right) with $\kappa t = 0.03$ followed by 10 applications of the ST (red), sBs (blue), and BsB (green) protocols. Here we take $\bar{n}=10$.}
\label{figSM:fidelity_vs_sqlevel2}
\end{figure}

In Fig.~\ref{figSM:fidelity_vs_sqlevel2}, we plot the entanglement fidelity $F_e$ against the squeezing parameter $r$.
We see that the entanglement fidelity indeed recovers by applying the QEC circuit. 
For smaller values of the squeezing parameter $r$, the entanglement fidelity is worse, since the QEC circuit dissipates the state into a manifold different from the SC state manifold.


\subsection{Logical operations with dissipation}
\subsubsection{$Z(\theta)$ operation}
\begin{figure}[htbp]
    \centering
    \includegraphics[width=0.5\linewidth]{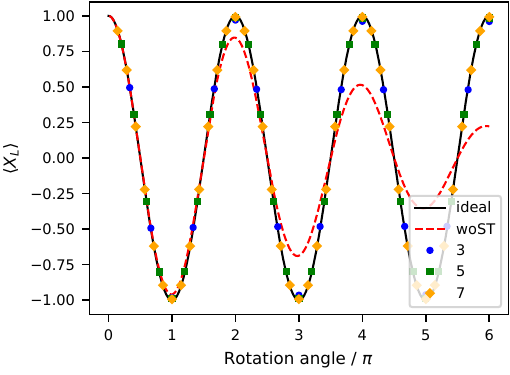}
    \caption{$Z$-rotation gate. The expectation value of $\h X_L$ after applying the displacement operator $\h D(i\theta/4\alpha)$ to $\ket{+}_L\ket 0_G$ state. The displacement operator drives the state away from the logical space, so the expectation value of $\h X_L$ exhibits a damped oscillation without STs. The application of STs dissipates the state back to the logical space, suppressing the decay of the oscillation amplitude. Parameters are set to be $\alpha = 2$ and $r = 1$.}
    \label{figSM:Zrot}
\end{figure}

Next, we numerically verify that the displacement operation given in Eq.~\eqref{eq:logx} induces the logical $Z$ rotation. To see this, we prepare the initial state as $\ket{+}_L\ket 0_G$ and then apply the displacement operator $\h D(i\theta/4\alpha)$. We also apply the QEC circuit periodically during the application of the displacement.
In Fig.~\ref{figSM:Zrot}, the expectation value of the logical $X$ operator is plotted. Without the QEC circuits, the amplitude of the oscillation decays due to the excitations in the gauge mode. However, by applying the QEC circuit pair $N$ times per $\pi$-rotation, the gauge excitations are removed and the decay is suppressed.

\subsubsection{$ZZ(\theta)$ operation}
\begin{figure}[htbp]
    \centering
    \includegraphics[width=0.5\linewidth]{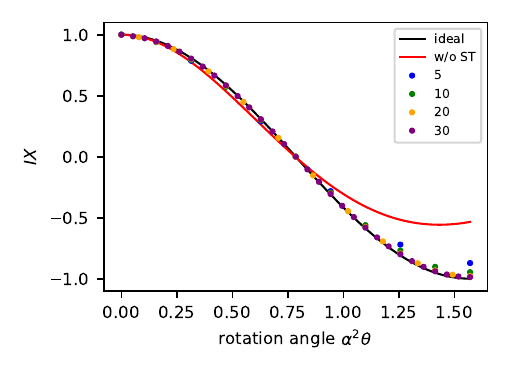}
    \caption{$ZZ$-rotation gate. The expectation value of $I_L\otimes X_L$ after the beam splitter interaction plotted against the rotation angle $\alpha^2\theta$. Parameters are set to be $\alpha = 2$ and $r = 1$.}
    \label{figSM:ZZrot}
\end{figure}

We numerically verify that the beam-splitter interaction given in Eq.~\eqref{eq:logx} induces the logical $ZZ$ rotation. To see this, we prepare the initial state as $\ket{+}_L\ket 0_G$ and then apply the displacement operator $\h D(i\theta/4\alpha)$. We also apply the QEC circuit periodically during the application of the displacement.
In Fig.~\ref{figSM:ZZrot}, the expectation value of the logical $X$ operator is plotted. Without the QEC circuits, the amplitude of the oscillation decays due to the excitations in the gauge mode. However, by applying the QEC circuit pair $N$ times per $\pi$-rotation, the gauge excitations are removed and the decay is suppressed.

\subsection{Logical measurement}
Finally, we numerically confirm the effectiveness of our improved circuit for measuring $\h Z_L$.
As possible realizations of the measurement of $Z_L$, we consider the na{\"i}ve Hadamard test for $Z_0$, using measurement circuits corresponding to several types of Trotterization of Eq.~\eqref{eq:MeasUnit} (sharpen, trim, BsB, and sBs), and the Homodyne measurement.
We define the error probability as $p_{\rm err}:=(p(1|0)+p(0|1))/2$, where $p(1|0) (p(0|1))$ is the probability of obtaining the measurement outcome 1(0) where the true state is $\ket 0_L (\ket 1_L)$.

\begin{figure}[htbp]
    \centering
    \includegraphics[width=0.5\linewidth]{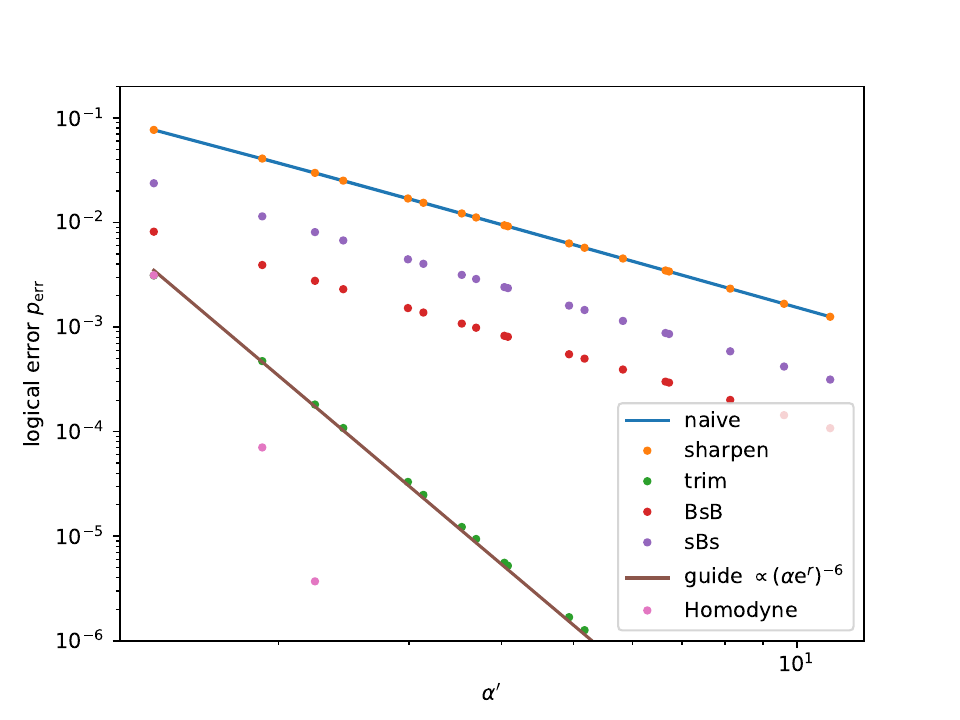}
    \caption{Measurement error $p_{\rm err} = (p(1|0)+p(0|1))/2$ in the logical measurement of $\h{Z}_L$ with different protocols. For the na{\"i}ve protocol, the logical error scales as $p_{\rm err}\propto \alpha'^{-2}$, while $p_{\rm err}\propto \alpha'^{-6}$ for the trim-like circuit in Fig.~\ref{fig:LogZ circuit}. }
    \label{figSM:LogicalMeasurement}
\end{figure}

In Fig.~\ref{figSM:LogicalMeasurement}, we plot the error probability for different measurement protocols against the rescaled displacement $\alpha'=\alpha e^r$.
We confirm that the error probability in the trim circuit measurement scales as $\alpha'^{-6}$, while that for other circuit-based protocols scales as $\alpha'^{-2}$, thereby showing the cubic improvement.

\end{document}